\documentclass[superscriptaddress,groupedaddress,nofootnoteinbib,12pt]{article} 
\usepackage{graphicx}
\usepackage{color}
\usepackage{dcolumn}   
\usepackage{bm}     
\usepackage{bbm}       
\usepackage{amssymb}  
\usepackage{amsmath}
\usepackage{sectsty}
\usepackage{latexsym}
\usepackage{float}
\usepackage{ifthen}
\usepackage{caption,subfig}
\usepackage{enumerate}
\usepackage{url}
\usepackage{caption,subfig}
\usepackage{jcappub}
\usepackage{amsopn}

\usepackage{amsfonts}
\usepackage{multirow}
\usepackage{array}
\usepackage{booktabs}
\usepackage{rotating}

\def\clap#1{\hbox to 0pt{\hss#1\hss}}

\def\({\left(}
\def\){\right)}
\def\[{\left[}
\def\]{\right]}
\def\bea{\begin{eqnarray}}
\def\eea{\end{eqnarray}}
\def\be{\begin{equation}}
\def\ee{\end{equation}}

\def\d{\mathrm{d}}

\newcommand{\Lag}{\mathcal{L}}
\newcommand{\Ham}{\mathcal{H}}

\newcommand{\Msc}{M_{\rm sc}}
\newcommand{\Msd}{M_{\rm sd}}
\newcommand{\Mvc}{M_{\rm vc}}
\newcommand{\Mvd}{M_{\rm vd}}
\newcommand{\MF}{M_F}

\newcommand{\Ei}{{\rm Ei}}

\def\clap#1{\hbox to 0pt{\hss#1\hss}}

\renewcommand{\geq}{\geqslant}

\newcommand{\mK}{\mathcal{K}}
\newcommand{\mU}{\mathcal{U}}
\newcommand{\mS}{\mathcal{S}}
\newcommand{\mG}{\mathcal{G}}
\newcommand{\Od}{\mathcal{O}}

\newcommand{\divA}{\partial\cdot A}

\definecolor{forestgreen}{rgb}{0.133,0.545,0.133}
\newboolean{editorial}
\setboolean{editorial}{true}
\newcommand{\editorial}[2]{\ifthenelse{\boolean{editorial}}{\textcolor{red}{[\textsf{\textbf{{#1}}}: }\textcolor{blue}{\textsf{{#2}}}\textcolor{red}{]}}{}}

\renewcommand{\d}{\mathrm{d}}

 \def\be   {\begin{equation}}   \def\ee   {\end{equation}}

 \def\ba  {\begin{eqnarray}}   \def\ea  {\end{eqnarray}}

\begin{document}

\title{On scalar and vector fields coupled to the energy-momentum tensor}

\author{Jose Beltr\'an Jim\'enez$^{a,b}$, Jose A. R. Cembranos$^c$ and Jose M. S\'anchez Vel\'azquez$^c$}
\affiliation{$^a$Instituto de F\'isica Te\'orica UAM-CSIC, Universidad Aut\'onoma de Madrid, Cantoblanco, Madrid, E-28049, Spain.\\
$^b$Departamento de F\'isica Fundamental, Universidad de Salamanca, E-37008 Salamanca, Spain.\\
$^c$Departamento de F\'isica Te\'orica and UPARCOS,
Universidad Complutense de Madrid, E-28040 Madrid, Spain.}

	\emailAdd{jose.beltran@uam.es}
	\emailAdd{cembra@fis.ucm.es}
	\emailAdd{jmsvelazquez@ucm.es}

\abstract{We consider theories for scalar and vector fields coupled to the energy-momentum tensor. Since these fields also carry a non-trivial energy-momentum tensor, the coupling prescription generates self-interactions. In analogy with gravity theories, we build the action by means of an iterative process that leads to an infinite series, which can be resumed as the solution of a set of differential equations. We show that, in some particular cases, the equations become algebraic and that is also possible to find solutions in the form of polynomials. We briefly review the case of the scalar field that has already been studied in the literature and extend the analysis to the case of derivative (disformal) couplings. We then explore theories with vector fields, distinguishing between gauge- and non-gauge-invariant couplings. Interactions with matter are also considered, taking a scalar field as a proxy for the matter sector. We also discuss the ambiguity introduced by superpotential (boundary) terms in the definition of the energy-momentum tensor and use them to show that it is also possible to generate Galileon-like interactions with this procedure. We finally use collider and astrophysical observations to set constraints on the dimensionful coupling which characterises the phenomenology of these models.}

IFT-UAM/CSIC-18-031

\maketitle

\section{Introduction}

General Relativity (GR) is the standard framework to describe the gravitational interaction and, after more than a century since its inception, it still stands out as the most compelling candidate owed to its excellent agreement with observations on a wide regime of scales \cite{Will:2014kxa}. From a theoretical viewpoint, GR can be regarded as the theory describing an interaction mediated by a massless spin 2 particle. The very masslessness of this particle together with explicit Lorentz invariance makes it to naturally couple to the energy-momentum tensor and, since it also carries energy-momentum, consistency dictates that it needs to present self-interactions. This requirement has sometimes led to regard gravity as a theory for a spin 2 particle that is consistently coupled to its own energy-momentum tensor so that the total energy-momentum tensor is the source of the gravitational field\footnote{Let us note however that the actual crucial requirement is to maintain the gauge symmetry at the non-linear level.}. These interactions can be constructed order by order following the usual Noether procedure (see for instance \cite{Ortin}) and one obtains an infinite series of terms. One could attempt to re-sum the series directly or to use Deser's procedure \cite{Deser:1969wk} of introducing auxiliary fields so that the construction of the interactions ends at the first iteration. Either way, GR arises as the full non-linear theory and the equivalence principle together with diffeomorphism symmetry come along in a natural way (see also \cite{Padmanabhan:2004xk,Butcher:2009ta,Deser:2009fq,Barcelo:2014mua} for some recent related works discussing in detail the bootstrapping procedure). 

In this work, we intend to develop a family of theories for scalar and vector fields following a similar bootstrapping approach as the one leading to GR, i.e., by prescribing a coupling to the energy-momentum tensor that remains at the full non-linear level. Unlike the case of gravity where the coupling to the energy-momentum tensor comes motivated from the requirement of maintaining gauge invariance (so it is a true consistency requirement rather than a prescription), in our case there is no {\it necessity} to have a {\it consistent} coupling to the energy-momentum tensor nor self-couplings of this form. However, the construction of theories whose interactions are universally described in terms of the energy-momentum tensor (as to fulfill some form of equivalence principle) is an alluring question in relation with gravitational phenomena. Let us remind that, starting from Newton's law, the simplest (perhaps naive) relativistic completion is to promote the gravitational potential to a scalar field. However, the most leading order coupling of the scalar to the energy-momentum tensor is through its trace and, therefore, there is no bending of light. This is a major obstacle for this simple theory of gravity based on a scalar field since the bending of light is a paramount feature of the gravitational interaction. Nevertheless, the problem of finding a theory for a scalar field that couples in a self-consistent manner to the energy-momentum tensor is an interesting problem on its own that has already been considered in the literature \cite{Kraichnan:1955zz,Freund:1969hh,Deser:1970zzb,Sami:2002se}. Here we will extend those results for the case of more general couplings for a scalar field (adding a shift symmetry that leads to derivative couplings) and explore the case of vector fields coupled to the energy-momentum tensor.

The paper is organised as follows: We start by briefly reviewing and re-obtaining known results for a scalar field coupled to the trace of the energy-momentum tensor. We then extend the results to incorporate a shift-symmetry for the scalar in the coupling to the energy-momentum tensor, what leads to a theory for a derivatively coupled scalar. After obtaining results in the second order formalism, we turn to discuss the construction of the full non-linear theories in the first order formalism, where the resummation procedures can be simplified. We end the scalar field case by considering couplings to matter. After working out the scalar field case, we consider theories for a vector field coupled to the energy-momentum tensor. We will devote Sec. \ref{sec:Superpotentials} to discuss the role played by superpotential terms and Sec. \ref{Sec:EffectiveMetrics} to present a procedure to obtain the interactions from a generating functional defined in terms of an effective metric. In Sec. \ref{Sec:Phenomenology} we give constraints obtained from several phenomenological probes and we conclude in Sec. \ref{Sec:Discussion} with a discussion of our results.

\section{Scalar gravity}
\label{Sec:ScalarGravity}
We will start our study with the simplest case of a scalar field theory that couples to the trace of the energy-momentum tensor to recover known results for scalar gravity. Then, we will extend these results to include derivative interactions that typically arise from disformal couplings. Such couplings will be the natural ones when imposing a shift symmetry for the scalar field, as usually happens for Goldstone bosons. We will also consider the problem from a first order point of view. Finally, couplings to matter, both derivative and non-derivative, will be constructed.

\subsection{Self-interactions for scalar gravity}\label{sec:Conformalscalar}
Let us begin our tour on theories coupled to the energy-momentum tensor from a scalar field and focus on the self-coupling problem neglecting other fields, i.e., we will look for consistent couplings of the scalar field to its own energy-momentum tensor. Firstly, we need to properly define our procedure. Our starting point will be the action for a free massive scalar field given by
\be
\mS_{(0)}=\frac12\int\d^4x\left(\partial_\mu\varphi\partial^\mu\varphi-m^2\varphi^2 \right).
\ee
The goal now is to add self-interactions of the scalar field  through couplings to the energy-momentum tensor. This can be done in two ways, either by imposing a coupling of the scalar field to its own energy-momentum tensor at the level of the action or by imposing its energy-momentum tensor to be the source in the field equations. We will solve both cases for completeness and to show the important differences that arise in both procedures at the non-linear level. Let us start by adding an interaction of the scalar to the energy-momentum tensor of the free field in the action as follows
\be
\mS_{(1)}=-\frac{1}{\Msc}\int\d^4x\;\varphi T_{(0)}
\label{eq:Firstcouplingphi}
\ee
with $\Msc$ some mass scale determining the strength of the interaction and $T_{(0)}$ the trace of the energy-momentum tensor of the free scalar field, i.e., the one associated to $\mS_{(0)}$. We encounter here the usual ambiguity due to the different available definitions for the energy-momentum tensor that differ either by a term of the form $\partial_\alpha \Theta^{[\alpha\mu]\nu}$ with $\Theta^{[\alpha\mu]\nu}$ some super-potential antisymmetric in the first pair of indices so that it is off-shell divergenceless, or by a term proportional to the field equations (or more generally, any rank-2 tensor whose divergence vanishes on-shell). In both cases, the form of the added piece guarantees that all of the related energy-momentum tensors give the same Lorentz generators, i.e., they carry the same total energy and momentum. We will consider in more detail the role of such boundary terms in Sec. \ref{sec:Superpotentials} and, until then, we will adopt the Hilbert prescription to compute the energy-momentum tensor in terms of a functional derivative with respect to an auxiliary metric tensor as follows\footnote{This definition is not free from subtleties either since one also needs to choose a covariantisation procedure and the tensorial character of the fields. We will assume a minimal coupling prescription for the covariantisation and that all the fields keep their tensorial character as in the original Minkowski space.}
\be
T^{\mu\nu}\equiv\left(-\frac{2}{\sqrt{-\gamma}}\frac{\delta \mS[\gamma_{\mu\nu}]}{\delta \gamma_{\mu\nu}}\right)_{\gamma_{\mu\nu}=\eta_{\mu\nu}},
\ee
where in the action we need to replace $\eta_{\mu\nu}\to \gamma_{\mu\nu}$ with $\gamma_{\mu\nu}$ some background (Lorentzian) metric and $\gamma$ its determinant. This definition has the advantage of directly providing a symmetric and gauge-invariant (in case of fields with spin and/or internal gauge symmetries) energy-momentum tensor. In general, this does not happen for the canonical energy-momentum tensor obtained from Noether's theorem, although the Belinfante-Rosenfeld procedure \cite{BelinfanteRosenfeld} allows to {\it correct} it and transform it into one with the desired properties \footnote{See also \cite{Gotay&Marsden} for a method to construct an energy-momentum tensor that can be interpreted as a generalisation of the Belinfante-Rosenfeld procedure.}. For the scalar field theory we are considering, the energy-momentum tensor is
\be
T_{(0)}^{\mu\nu}=\partial^\mu\varphi\partial^\nu\varphi-\eta^{\mu\nu}\Lag_\varphi,
\ee
which is also the one obtained as Noether current so that the above discussion is not relevant here. However, in the subsequent sections dealing with vector fields this will be important since the canonical and Hilbert energy-momentum tensors differ. 

After settling the ambiguity in the energy-momentum tensor, we can now write the first order corrected action for $\mS_{(0)}$ by incorporating the coupling (\ref{eq:Firstcouplingphi}), so we obtain
\be
\mS_{(0)}+\mS_{(1)}=\frac12\int\d^4x\left[\left(1+2\frac{\varphi}{\Msc}\right)\partial_\mu\varphi\partial^\mu\varphi-\left(1+4\frac{\varphi}{\Msc}\right)m^2\varphi^2\right].
\ee
As usual, when we introduce the coupling of the scalar field to the energy-momentum tensor, the new energy-momentum tensor of the whole action acquires a new contribution and, therefore, the coupling $-\frac{\varphi}{M_{sc}}T$ receives additional corrections that will contribute to order $1/\Msc^2$. The added $1/\Msc^2$ interaction will again add a new correction that will contribute an order $1/\Msc^3$ term and so on. This iterative process will continue indefinitely so  we end up with a construction of the interactions as a perturbative expansion in powers of $\varphi/\Msc$ and, thus, we obtain an infinite series whose resummation will give the final desired action. The iterative process for the case at hand gives the following expansion for the first few terms:
\bea
\mS=\frac12\int\d^4x\left[\left(1+2\frac{\varphi}{\Msc}+4\frac{\varphi^2}{\Msc^2}+8\frac{\varphi^3}{\Msc^3}+\cdots\right)\partial_\mu\varphi\partial^\mu\varphi\right.\nonumber\\
\left.-\left(1+4\frac{\varphi}{\Msc}+16\frac{\varphi^2}{\Msc^2}+64\frac{\varphi^3}{\Msc^3}+\cdots\right)m^2\varphi^2\right].
\label{eq:PertScalar1}
\eea
It is not difficult to identify that we obtain the first terms of a geometric progression with ratios $2\varphi/\Msc$ and $4\varphi/\Msc$ which can then be easily resummed. One can confirm this by realising that a term $(\varphi/\Msc)^n\partial_\mu\varphi\partial^\mu\varphi$ gives a correction $2(\varphi/\Msc)^{n+1}\partial_\mu\varphi\partial^\mu\varphi$, while a term $(\varphi/\Msc)^{m+2}$ introduces a correction $4(\varphi/\Msc)^{m+3}$. Then, the resummed series will be given by
\bea
\mS=\frac12\int\d^4x\Big[\mK(\varphi)\partial_\mu\varphi\partial^\mu\varphi-\mU(\varphi)m^2\varphi^2\Big],
\label{Actionscalar1}
\eea
with
\bea
\mK(\varphi)=\sum_{n=0}^\infty\left(2\frac{\varphi}{M}\right)^n=\frac{1}{1-2\varphi/\Msc},\quad\quad \mU(\varphi)=\sum_{n=0}^\infty\left(4\frac{\varphi}{M}\right)^n=\frac{1}{1-4\varphi/\Msc}.
\label{resconformalscalar}
\eea
This recovers the results in \cite{Sami:2002se} in the corresponding limits. Technically, the geometric series only converges for\footnote{The convergence of the generated perturbative series will be a recurrent issue throughout this work. In fact, most of the obtained series will need to be interpreted as asymptotic series of the true underlying theory. We will discuss this issue in more detail in due time.} $2\varphi<\Msc$, but the final result can be extended to values  $2\varphi>\Msc$, barring the potential poles at $2\varphi/\Msc=1$ and $4\varphi/\Msc=1$ that occur for positive values of the scalar field, assuming $\Msc>0$, while for $\varphi<0$ the functions are analytic. Let us also notice that, had we started with an arbitrary potential for the scalar field $V(\varphi)$ instead of a mass term, the corresponding final action would have resulted in a re-dressed potential with the same factor, i.e., the effect of the interactions on the potential would be $V(\varphi)\to \mU(\varphi) V(\varphi)$ and, as a particular case, if we start with a constant potential $V_0$ corresponding to a cosmological constant, the same re-dressing will take place so that the cosmological constant becomes a $\varphi-$dependent quantity. In any case, we find it more natural to start with a mass term in compliance with the prescribed procedure of generating the interactions through the coupling to the energy-momentum tensor, i.e., the natural starting point is the free theory.

An alternative way to resum the series that will be very useful in the less obvious cases that we will consider later is to notice that the resulting perturbative expansion (\ref{eq:PertScalar1}) allows to guess the final form of the action to be of the form (\ref{Actionscalar1}). Then, we can impose the desired form of our interactions to the energy-momentum tensor so that the full non-linear action must satisfy
\bea
\mS=\int\d^4x\left(\frac12\mK(\varphi)\partial_\mu\varphi\partial^\mu\varphi-\mU(\varphi)V(\varphi)\right)=\int\d^4x\left(\frac12\partial_\mu\varphi\partial^\mu\varphi-V(\varphi)-\frac{\varphi}{\Msc}T\right),
\label{Actionscalar2}
\eea
with $T$ the trace of the energy-momentum tensor of the full action, i.e.,
\be
T=-\mK(\varphi)\partial_\mu\varphi\partial^\mu\varphi+4\mU V.
\ee
We have also included here an arbitrary potential for generality. Thus, we will need to have
\be
\mS=\int\d^4x\left(\frac12\mK\partial_\mu\varphi\partial^\mu\varphi-\mU V\right)=\int\d^4x\left[\frac12\Big(1+\frac{2\varphi}{\Msc}\mK\Big)\partial_\mu\varphi\partial^\mu\varphi-\Big(1+\frac{4\varphi}{\Msc}\mU\Big)V\right]
\label{Actionscalar3}
\ee
from which we can recover the solutions for $\mK$ and $\mU$ given in (\ref{resconformalscalar}). Notice that this method allows to obtain the final action without relying on the convergence of the perturbative series and, thus, the aforementioned extension of the resummed series is justified. As a final remark, it is not difficult to see that, had we started with a coupling to an arbitrary function of $\varphi$ of the form $-f(\varphi/\Msc)T_{(0)}$, the final result would be the same with the replacement $\varphi/\Msc\to f(\varphi/\Msc)$ in the final form of the function $\mK$ and $\mU$, recovering that way the results of \cite{Sami:2002se}. 

We have then obtained the action for a scalar field coupled to its own energy-momentum tensor at the level of the action. However, as we mentioned above, we can alternatively impose the trace of the energy-momentum tensor to be the source of the scalar field equations, i.e., the full theory must lead to equations of motion satisfying
\be
(\Box+m^2)\varphi=-\frac{1}{\Msc} T,
\label{equationscalar1}
\ee
again with $T$ the total energy-momentum tensor of the scalar field. Before proceeding to solve this case, let us comment on some important differences with respect to the gravitational case involving a spin-2 field. The above equation is perfectly consistent at first order, i.e., we could simply add $T_{(0)}$ on the RHS, so we already have a consistent theory and there is no need to include higher order corrections. This is in high contrast with the construction in standard gravity where the Bianchi identities for the spin-2 field (consequence of the required gauge symmetry) are incompatible with the conservation of the energy-momentum tensor and one must add higher order corrections to have consistent equations of motion. For the scalar gravity case, although not imposed by the consistency of the equations, we can extend the construction in an analogous manner and impose that the source of the equation is not given in terms of the energy-momentum tensor of the free scalar field, but the total energy-momentum tensor. As before, we could proceed order by order to find the interactions, but we will directly resort to guess the final action to be of the form given in (\ref{Actionscalar1}) and obtain the required form of the functions $\mK$ and $\mU$ for the field equations to be of the form given in (\ref{equationscalar1}). For the sake of generality, we will consider a general bare potential $V(\varphi)$ instead of a simple mass term. By varying (\ref{Actionscalar1}) w.r.t. the scalar field we obtain
\be
\Box\varphi=-\frac{\mK'}{2\mK}(\partial\varphi)^2-\frac{(\mU V)'}{\mK}
\ee
that must be compared with the prescribed form of the field equation
\be
\Box\varphi+V'=-\frac{1}{\Msc} T=\frac{1}{\Msc}\Big[\mK(\partial\varphi)^2-4\mU V\Big].
\label{equationscalar}
\ee
Thus, we see that the functions $\mK$ and $\mU$ must satisfy the following equations
\be
\mK'=-\frac{2\mK^2}{\Msc},\quad\quad\mU'+\left(\frac{V'}{V}-\frac{4\mK}{\Msc}\right)\mU=\mK\frac{V'}{V}.
\ee
The solution for $\mK$ can be straightforwardly obtained to be
\be
\mK=\frac{1}{1+2\varphi/\Msc}
\label{eq:solKsource}
\ee
where we have chosen the integration constant so that $\mK(0)=1$, i.e., we absorbed $\mK(0)$ into the normalization of the free field. It might look surprising that the solution for $\mK$ in this case is related to (\ref{resconformalscalar}) by a change of sign of $\varphi$. This could have been anticipated by noticing that the construction of the theory so that $T$ appears as a source of the field equations requires an extra minus sign with respect to the coupling at the level of the action to compensate for the one introduced by varying the action. Thus, the two series only differ by this extra $(-1)^n$ factor in the series that results in the overall change of sign of $\varphi$. 

From the obtained equations we see that the solution for $\mU$ depends on the form of the bare potential $V(\varphi)$. We can solve the equation for an arbitrary potential and the solution is given by
\be
\mU=\frac{(1+2\varphi/\Msc)^2}{V(\varphi)}\left[C_1+\int\frac{V'(\varphi)}{(1+2\varphi/\Msc)^3}\d\varphi\right]
\ee
with $C_1$ an integration constant that must be chosen so that $\mU(0)=1$. If $V'(\varphi)\neq0$, we need to set $C_1=0$. Remarkably, if we take a quadratic bare potential corresponding to adding a mass for the scalar field (which is the most natural choice if we start from a free theory), the above solution reduces to $\mU=1$. In that case, the resummed action reads
\be
\mS=\frac{1}{2}\int\d^4x\left[\frac{(\partial\varphi)^2}{1+2\varphi/\Msc}-m^2\varphi^2\right],
\ee
which is the action already obtained by Freund and Nambu in \cite{Freund:1969hh}, and which reduces to Nordstr\o m's theory in the massless limit. We have obtained here the solution for the more general case with an arbitrary bare potential, in which case the solution for $\mU$ depends on the form of the potential. Finally, if we have $V'=0$, the starting action contains a cosmological constant $V_0$ and the obtained solution for $\mU(\varphi)$ gives the scalar field re-dressing of the cosmological constant, which is $(1+2\varphi/\Msc)^2$ obtained after setting $C_1=V_0$ as it corresponds to have $\mU(0)=1$. We will re-obtain this result in Sec. \ref{Sec:Scalarmattercoupling} when studying couplings to matter fields.

\subsection{Derivatively coupled scalar gravity}
\label{Sec:ScalarDerivative}
After warming up with the simplest coupling of the scalar field to the trace of its own energy-momentum tensor, we will now look at interactions enjoying a shift symmetry, what happens for instance in models where the scalar arises as a Goldstone boson, a paradigmatic case in gravity theories being branons, that are associated to the breaking of translations in extra dimensions \cite{DoMa}. This additional symmetry imposes that the scalar field must couple derivatively to the energy-momentum tensor and this further imposes that the leading order interaction must be quadratic in the scalar field, i.e., we will have a coupling of the form $\partial_\mu\varphi\partial_\nu\varphi T^{\mu\nu}$. This is also the interaction arising in theories with  disformal couplings \cite{Disformal}. Although an exact shift symmetry is only compatible with a massless scalar field, we will leave a mass term for the sake of generality (and which could arise from a softly breaking of the shift symmetry). In fact, again and for the sake of generality, we will consider a general bare potential term. Then, the action with the first order correction arising from the derivative coupling to the energy-momentum tensor in this case is given by
\be
\mS_{(0)}+\mS_{(1)}=\int\d^4x\left[\frac12\partial_\mu\varphi\partial^\mu\varphi-V(\varphi)+\frac{1}{\Msd^4}\partial_\mu\varphi\partial_\nu\varphi T^{\mu\nu}_{(0)}\right],
\ee
with $\Msd$ some mass scale. It is interesting to notice that now the coupling is suppressed by $\Msd^{-4}$ so that the leading order interaction corresponds to a dimension 8 operator, unlike in the previous non-derivative coupling whose leading order was a dimension 5 operator. As before, the added interaction will contribute to the energy-momentum tensor so that the interaction needs to be corrected. If we proceed with this iterative process, we find the expansion
\begin{align}
\mS_\varphi=&\int\d^4x\left[\frac12\Big(1+X+3X^2+15X^3+\cdots \Big)\partial_\mu\varphi\partial^\mu\varphi-\Big(1-X-X^2-3X^3 \cdots \Big)V(\varphi) \right].
\label{eq:Pert1}
\end{align}
where we have defined $X\equiv(\partial\varphi)^2/\Msd^4$. Again, we could obtain the general term of the generated series and eventually resum it. However, it is easier to use an Ansatz for the resummed action by noticing that, from (\ref{eq:Pert1}), we can guess the final form of the action to be
\be
\mS_\varphi=\int\d^4x\left[\frac12\mK(X)\partial_\mu\varphi\partial^\mu\varphi-\mU(X)V(\varphi) \right]
\label{eq:AnsatzScalar}
\ee
with $\mK(X)$ and $\mU(X)$ some functions to be determined from our prescribed couplings. Thus, by imposing that the final action must satisfy
\be
\mS_\varphi=\int\d^4x\left[\frac12\partial_\mu\varphi\partial^\mu\varphi-V(\varphi) +\frac{1}{\Msd^4}\partial_\mu\varphi\partial_\nu\varphi T^{\mu\nu}\right]
\ee
with $T_{\mu\nu}$ the total energy-momentum tensor, we obtain the following relation:
\bea
\mS_\varphi&=&\int\d^4x\left[\frac12\mK(X)\partial_\mu\varphi\partial^\mu\varphi-\mU(X)V(\varphi) \right]\\
&=&\int\d^4x\left[\frac12\Big(1+X\mK(X)+2X^2\mK'(X)  \Big)\partial_\mu\varphi\partial^\mu\varphi-\Big(1-X\mU(X)+2X^2\mU'(X)  \Big)V(\varphi) \right],\nonumber
\eea
where the prime stands for derivative w.r.t. its argument. Thus, the functions $\mK(X)$ and $\mU(X)$ will be determined by the following first order differential equations
\bea
\mK(X)&=&1+X\mK(X)+2X^2\mK'(X),\\
\mU(X)&=&1-X\mU(X)+2X^2\mU'(X).
\label{eq:DerivativeScalar}
\eea
We have thus reduced the problem of resuming the series to solving the above differential equations. The existence of solutions for these differential equations will guarantee the convergence (as well as the possible analytic extensions) of the perturbative series. Although not important for us here, it is possible to obtain the explicit analytic solutions as
\bea
\mK&=&-\frac{e^{-1/(2X)}}{2X}\Ei_{1/2}\big(-1/(2X)\big)\\
\mU&=&-\frac{e^{-1/(2X)}}{2X}\Ei_{-1/2}(-1/(2X)).
\eea
where $\Ei_n(x)$ stands for the exponential integral function of order $n$ and we have chosen the integration constants in order to have a well-defined solution for $X\to0$. In principle, one might think that boundary conditions must be imposed so that $\mK(0)=\mU(0)=1$. However, these boundary conditions are actually satisfied by all solutions of the above equations since they are hardwired in the own definition of the functions $\mK$ and $\mU$ through the perturbative series. The way to select the right solution is thus by imposing regularity at the origin $X=0$. Even this condition is not sufficient to select one single solution and this is related to the fact that the perturbative series must be interpreted as an asymptotic expansion\footnote{In fact, the solution resembles one of the paradigmatic examples of asymptotic expansion $e^{-1/t}\Ei(1/t)=\sum_{n=0}^\infty n! t^{n+1}$.}, rather than a proper series expansion. In fact, it is not difficult to check that the perturbative series is divergent, as it is expected for asymptotic expansions. Thus, the above solution is actually one of many different possible solutions. We will find these equations often and we will defer a more detailed discussion of some of their features to the Appendix \ref{Appendix}.

So far we have focused on the coupling $\partial_\mu\varphi \partial_\nu \varphi T^{\mu\nu}$, but, at this order, we can be more general and allow for another interaction of the same dimension so that the first correction becomes
\be
\mS_{(1)}=\frac{1}{\Msd^4}\int\d^4x\Big(b_1\partial_\mu\varphi\partial_\nu\varphi+b_2\partial_\alpha\varphi\partial^\alpha\varphi \eta_{\mu\nu}\Big) T^{\mu\nu}_{(0)},
\ee
where $b_1$ and $b_2$ are two arbitrary dimensionless parameters, one of which could actually be absorbed into $\Msd$, but we prefer to leave it explicitly to keep track of the two different interactions. The previous case then reduces to $b_2=0$, which is special in that the coupling does not depend on the metric and, as we will see in Sec. \ref{Sec:EffectiveMetrics}, this has interesting consequences in some constructions. For this more general coupling, the perturbative series is
\begin{align}
\mS_\varphi=&\int\d^4x\left[\frac12\Big(1+(b_1-2b_2)X+3b_1(b_1-2b_2)X^2+3b_1(b_1-2b_2)(5b_1+2b_2)X^3+\cdots \Big)\partial_\mu\varphi\partial^\mu\varphi\right.\nonumber\\
&\left. -\Big(1-(b_1+4b_2)\left(X+(b_1-2b_2)X^2+3b_1(b_1-2b_2)X^3 \cdots \right)\Big)V(\varphi) \right].
\label{eq:Pert2}
\end{align}
To resum the series we can follow the same procedure as before using the same Ansatz for the resummed action as in (\ref{eq:AnsatzScalar}), in which case we obtain that the following relation must hold:
\bea
\mS&=&\int\d^4x\left[\frac12\mK(X)\partial_\mu\varphi\partial^\mu\varphi-\mU(X)V(\varphi) \right]\nonumber\\
&=&\int\d^4x\left[\frac12\partial_\mu\varphi\partial^\mu\varphi-V(\varphi)+\frac{1}{\Msd^4}\Big(b_1\partial_\mu\varphi\partial_\nu\varphi+b_2\partial_\alpha\varphi\partial^\alpha\varphi \eta_{\mu\nu}\Big) T^{\mu\nu}\right]\nonumber\\
&=&\int\d^4x\left[\frac12\Big(1+(b_1-2b_2)X\mK+2(b_1+b_2)X^2\mK'  \Big)\partial_\mu\varphi\partial^\mu\varphi\right.\nonumber\\
&&\hspace{1cm}\left.-\Big(1-(b_1+4b_2)X\mU+2(b_1+b_2)X^2\mU'  \Big)V(\varphi) \right].
\eea
Thus, the equations to be satisfied in this case are 
\bea
\mK&=&1+(b_1-2b_2)X\mK+2(b_1+b_2)X^2\mK',\\
\mU&=&1-(b_1+4b_2)X\mU+2(b_1+b_2)X^2\mU'.
\label{Eq:eqsderivativecoupling}
\eea
The additional freedom to choose the relation between the two free parameters $b_1$ and $b_2$ allows now to straightforwardly obtain some particularly interesting solutions. Firstly, for $b_1+b_2=0$, the equations become algebraic and the unique solution is given by
\bea
\mK(X)=\mU(X)=\frac{1}{1-3b_1X}.
\eea
This particular choice of parameters that make the equations algebraic is remarkable because it precisely corresponds to coupling the energy-momentum tensor to the orthogonal projector to the gradient of the scalar field $\eta_{\mu\nu}-\partial_\mu\varphi \partial_\nu\varphi/(\partial\varphi)^2$. On the other hand, we can see from the perturbative series (\ref{eq:Pert2}) that the condition $b_1-2b_2=0$ cancels all the corrections to the kinetic term and this can also be seen from the differential equations where it is apparent that, for those parameters, $\mK=1$ is the corresponding solution. Moreover, for that choice of parameters, we see from the perturbative expansion that $\mU=1-6b_2X$, which can be confirmed to be the solution of the equation for $\mU$ with $b_1=2b_2$. Likewise, for $b_1+4b_2=0$, all the corrections to the potential vanish and only the kinetic term is modified. It is worth mentioning that the iterative procedure used to construct the interactions also allows to obtain polynomial solutions of arbitrarily higher order by appropriately choosing the parameters. All these interesting possibilities are explained in more detail in the Appendix \ref{Appendix}.

Finally, let us notice that a constant potential $V_0$ that amounts to introducing a cosmological constant in the free action leads to a re-dressing of the cosmological constant analogous to what we found above for the non-derivative coupling, but with the crucial difference that now the cosmological constant becomes kinetically re-dressed in the full theory.

In the general case we see that we obtain a particular class of K-essence theories where the $\varphi$-dependence is entirely given by the starting potential, but it receives a kinetic-dependent re-dressing. On the other hand, if we start with an exact shift symmetry, given that the interactions do not break it, the resulting theory reduces to a particular class of $P(X)$ theories.

\subsection{First order formalism}
\label{Sec:ScalarFirstOrder}
In the previous section we have looked at the theory for a scalar field that is derivatively coupled to its own energy-momentum tensor. The problem was reduced to solving a couple of differential equations expressed in (\ref{eq:DerivativeScalar}). Here we will explore the same problem but from the first order formalism perspective. In the case of non-abelian gauge fields and also in the case of gravity, the first order formalism has proven to significantly simplify the problem since the iterative process ends at the first iteration \cite{Deser:1969wk}. For non-abelian theories, the first order formalism solves the self-coupling problem in one step instead of the four iterations required in the Lagrangian formalism. In the case of gravity, the simplification is even greater since it reduces the infinite iterations of the self-coupling problem to only one. This is in fact the route used by Deser to obtain the resummed action for the self-couplings of the graviton \cite{Deser:1969wk}. The significant simplifications in these cases encourage us to consider the construction of the theories with our prescription in the first order formalism in order to explore if analogous simplifications take place. As a matter of fact, the first order formalism for scalar gravitation was already explored in \cite{Deser:1970zzb} for the massless theory and with a conformal coupling so that the trace of the energy-momentum tensor appears as the source of the scalar field. It was then shown the equivalence of the resulting action with Nordstr\o m's theory of gravity and the massless limit of the theory obtained by Freund and Nambu \cite{Freund:1969hh} with the first order formalism (which we reproduced and extended above). We will use this formalism for the theories with derivative couplings to the energy-momentum tensor, what in the first order formalism means couplings to the canonical momentum.

The first thing we need to clarify is how we are going to define the theory in the first order formalism. The starting free theory for a massive scalar field $\varphi$ can be described by the following first order action:
\bea
\mS_{(0)}=\int\d^4x\left[\pi^\mu\partial_\mu\varphi-\frac12 \big(\pi^2+m^2\varphi^2)\right],
\eea
with $\pi^\mu$ the corresponding momentum in phase space. Upon variations with respect to the momentum $\pi^\mu$ and the scalar field we obtain the usual Hamilton equations $\partial_\mu\pi^\mu+m^2\varphi=0$ and $\pi_\mu=\partial_\mu\varphi$, which combined gives the desired equation $(\Box+m^2)\varphi=0$. At the lowest order we then prescribe a coupling to the energy-momentum tensor as\footnote{Of course, we could have also added a term $\pi^2 T$, but the considered coupling will be enough to show how the use of the first order formalism leads to simpler non-differential equations.}
\bea
\mS_{(0)}+\mS_{(1)}=\int\d^4x\left[\pi^\mu\partial_\mu\varphi-\frac12 \big(\pi^2+m^2\varphi^2\big)+\frac{1}{\Msd^4}\pi_\mu\pi_\nu T_{(0)}^{\mu\nu}\right]
\eea
with $T_{(0)}^{\mu\nu}$ the energy-momentum tensor corresponding to the free theory $\mS_{(0)}$. This is the form that we will also require for the final theory replacing $T_{(0)}^{\mu\nu}$ by the total energy-momentum tensor. Following the same reasoning as in the previous section, the final theory should admit an Ansatz of the following form:
\bea
\mS=\int\d^4x\left[\pi^\mu\partial_\mu\varphi-\Ham(\varphi,\pi^2)\right]
\eea
where the Hamiltonian\footnote{Let us stress that this Hamiltonian function will not give, in general, the energy of the system, although that is the case for homogeneous configurations.} $\Ham$ will be some function of the phase space coordinates. Lorentz invariance imposes that the momentum can only enter through its norm. As in the Lagrangian formalism, the energy-momentum tensor admits several definitions that differ by a super-potential term or quantities vanishing on-shell. As before, we shall resort to the Hilbert energy-momentum tensor. In this approach, one needs to specify the tensorial character of the fields, which are usually assumed to be true tensors. In some cases, it is however more convenient to assume that some fields actually transform as tensorial densities. In the Deser construction, assuming that the graviton is a tensorial density simplifies the computations. At the classical level and on-shell, assuming different weights only results in terms that vanish on-shell in the energy-momentum tensor\footnote{If we re-define a given field $\Phi$ with a metric-dependent change of variables $\Phi\rightarrow \Phi'=\Phi'(\Phi,\gamma_{\mu\nu})$ (as it happens when the re-definition corresponds to a change in the tensorial weight of the field), we have the following relation for the variation of the action
\be
\frac{\delta \mS[\Phi',\gamma_{\mu\nu}]}{\delta \gamma_ {\mu\nu}}=\left(\frac{\delta \mS[\Phi',\gamma_{\mu\nu}]}{\delta \gamma_ {\mu\nu}}\right)_{\Phi'}+\frac{\delta \mS}{\delta \Phi'} \frac{\delta \Phi'}{\delta \gamma_{\mu\nu}}.\nonumber
\ee
Since the second term on the RHS vanishes on the field equations of $\Phi'$ we obtain that both energy-momentum tensors coincide on-shell.\label{F7}}. In the present case, it is convenient to assume that $\pi^\mu$ is a tensorial density of weight 1 such that $p^\mu=\pi^\mu/\sqrt{-\gamma}$ is a tensor of zero weight. The advantage of using this variable is twofold: on one hand, this tensorial weight for $\pi^\mu$ makes $\pi^\mu\partial_\mu\varphi$ already a weight-0 scalar without the need to introduce the $\sqrt{-\gamma}$ in the volume element and, consequently, this term will not contribute to the energy-momentum tensor. On the other hand, the variation of the Hamiltonian $\Ham$ with respect to the auxiliary metric $\gamma_{\mu\nu}$ gives
\be
\delta\Ham(p^2)=\frac{\partial \Ham}{\partial p^2}\delta p^2=-\frac{\partial\Ham}{\partial\pi^2}\pi^2\(\gamma^{\mu\nu}-\frac{\pi^\mu\pi^\nu}{\pi^2}\)\delta \gamma_{\mu\nu}
\ee
which is proportional to the orthogonal projector to the momentum and, thus, although it does contribute to the energy-momentum tensor, it will not contribute to the interaction $T^{\mu\nu}\pi_\mu \pi_\nu$. To derive the above variation, we have taken into account that the Hamiltonian remains a scalar after the covariantisation and, because of the assumed weight of $\pi^\mu$, it will become a function of $\Ham(\varphi,\pi^2)\to\Ham\big(\varphi,p^2\big)=\Ham\big(\varphi,\pi^2/\vert\gamma\vert\big)$. The covariantised action then reads 
\be
\mS=\int\d^4x\left[\pi^\mu\partial_\mu\varphi-\sqrt{-\gamma}\Ham\big(\varphi,\pi^2/\vert\gamma\vert\big)\right]
\ee
and the total energy-momentum tensor computed with the described prescription is given by
\be
T^{\mu\nu}=-2\pi^2\frac{\partial\Ham}{\partial \pi^2}\left(\eta^{\mu\nu}-\frac{\pi^\mu\pi^\nu}{\pi^2}\right)+\Ham\eta^{\mu\nu}.
\label{Eq:Tfirstorder}
\ee
If we compare with the canonical energy-momentum tensor $\Theta^{\mu\nu}=2\frac{\partial\Ham}{\partial \pi^2}\pi^\mu\pi^\nu-\Lag\eta^{\mu\nu}$, we see that the difference is $\Ham-2\pi^2\frac{\partial\Ham}{\partial \pi^2}-\Lag$, which vanishes upon use of the Hamilton equation $\partial_\mu\varphi=\frac{\partial\Ham}{\partial \pi^\mu}$ and the relation between the Hamiltonian and the Lagrangian via a Legendre transformation $\Lag=\pi^\mu\partial_\mu\varphi-\Ham$. The final action must therefore satisfy the relation
\bea
\mS&=&\int\d^4x\left[\pi^\mu\partial_\mu\varphi-\Ham\right]
=\int\d^4x\left[\pi^\mu\partial_\mu\varphi-\frac12(\pi^2+m^2\varphi^2)+\frac{1}{\Msd^4}\pi_\mu\pi_\nu T^{\mu\nu}\right]\nonumber\\
&=&\int\d^4x\left[\pi^\mu\partial_\mu\varphi-\frac12(\pi^2+m^2\varphi^2)+\frac{1}{\Msd^4}\pi^2\Ham\right]
\eea
where we see that the chosen weight for $\pi^\mu$ has greatly simplified the interaction term that simply reduces to $\pi^2\Ham$. The above relation then leads to the algebraic equation
\be
\Ham=\frac12(\pi^2+m^2\varphi^2)-\frac{1}{\Msd^4}\pi^2\Ham
\ee
so that the Hamiltonian of the desired action will be given by
\be
\Ham=\frac12\frac{\pi^2+m^2\varphi^2}{1+\pi^2/\Msd^4}.
\ee
We see that the use of the first order formalism has substantially simplified the resolution of the problem since we do not encounter differential equations. Needless to say that the solutions obtained in both first and second order formalisms are different. This apparent ambiguity in the resulting theory as obtained with the first or the second order formalism actually reflects the ambiguity in the definition of the energy-momentum tensor because, as discussed above, the energy-momentum tensor Eq. (\ref{Eq:Tfirstorder}) differs from the one used in the second order formalism by a term that vanishes on-shell (see also Footnote \ref{F7}).  Expressing the theory obtained here in the second order formalism is not very illuminating so we will not give it, although it would be straightforward to do it. Let us finally notice that, if the leading order term in the Hamiltonian for the limit $\pi^2/\Msd^4\ll1$ is assumed to be $\Ham_0$, then it is not difficult to see that the full Hamiltonian will be
\be
\Ham=\frac{\Ham_0}{1+\pi^2/\Msd^4}.
\ee
i.e., the procedure simply re-dresses the seed Hamiltonian with the factor $(1+\pi^2/\Msd^4)^{-1}$. One interesting property of the resulting theory is that the Hamiltonian density $\Ham$ for the massless case saturates to the scale $\Msd^4$ at large momenta. 

\subsection{Coupling to matter fields}
\label{Sec:Scalarmattercoupling}
In the previous subsections we have found the action for the self-interacting scalar field through its own energy-momentum tensor, both with ultra-local and derivative couplings. Now we will turn our analysis to the couplings of the scalar field with other matter fields following the same philosophy, i.e., the scalar will couple to the energy-momentum tensor of matter fields. For simplicity, we will only consider the case of a matter sector described by a scalar field $\chi$. The derivative couplings will be the same as we will obtain for the vector field couplings to matter that will be treated in the next section so that, in order not to unnecessarily repeat the derivation, we will not give it here and discuss it in Sec. \ref{Sec:VectorMatter}. Thus, we will only deal with the conformal couplings so that our starting action for the proxy scalar field $\chi$ including the first order coupling to $\varphi$ is
\be
\mS_{\chi,(0)}+\mS_{\chi,(1)}=\int \d^4x\left[\frac12\partial_\mu\chi\partial^\mu\chi-W(\chi)-\frac{1}{\Msc}\varphi T_{\chi,(0)}\right]
\ee
where $T_{\chi,(0)}$ is the trace of the energy-momentum tensor of the proxy field and $W(\chi)$ the corresponding potential.
Going on in the iterative process yields the following series for the total action
\bea
\mS_{\chi}=\int \d^4 x\left[\frac12\left(1+\frac{2}{\Msc}\varphi+\frac{4}{\Msc^2}\varphi^2+ \frac{8}{\Msc^3}\varphi^3+...\right)\partial_\mu\chi\partial^\mu\chi\right.\nonumber\\ 
\left.-\left(1+\frac{4}{\Msc}\varphi+\frac{16}{\Msc^2}\varphi^2+\frac{64}{\Msc^3}\varphi^3+...\right)W(\chi)\right].
\eea
This expression has the form of a geometric series so it is straightforward to resum it yielding the final action for scalar gravity algebraically coupled to matter:
\be
\mS=\int \d^4x \left(\frac12\alpha(\varphi)\partial_\mu\chi\partial^\mu\chi-\beta(\varphi)W(\chi)\right)
\ee
with
\bea
\alpha(\varphi)=\sum_{n=0}^\infty\left(2\frac{\varphi}{M}\right)^n=\frac{1}{1-2\varphi/\Msc},\quad\quad \beta(\varphi)=\sum_{n=0}^\infty\left(4\frac{\varphi}{M}\right)^n=\frac{1}{1-4\varphi/\Msc}.
\eea
Not very surprisingly, we obtain the same result as for the self-couplings of the scalar field, i.e., both the kinetic and potential terms get re-dressed by the same factors as we found in Sec. \ref{sec:Conformalscalar} for $\varphi$. If the matter sector consists of a cosmological constant, which would correspond to a constant scalar field in the above solutions, the coupling procedure gives rise to an additional modification of the $\varphi$ potential or, equivalently, the cosmological constant becomes a $\varphi$-dependent quantity, which is also the result that we anticipated in Sec \ref{sec:Conformalscalar} for a constant potential of $\varphi$. Some phenomenological consequences of this mechanism were explored in a cosmological context in \cite{Sami:2002se}.

As in the self-coupling case, we could have introduced the coupling so that the trace of the energy-momentum tensor appears as a source of the scalar field equations. In order to obtain the theory with the required property, we shall follow the procedure of assuming the following action for the scalar gravity field $\varphi$ and the scalar proxy field $\chi$:
\bea
\mS=\int \d^4 x \left(\frac12\mK(\varphi)\partial_\mu\varphi\partial^\mu\varphi-\mU(\varphi)V(\varphi)+\frac12\tilde{\alpha}(\varphi)\partial_\mu\chi\partial^\mu\chi-\tilde{\beta}(\varphi)W(\chi)\right).
\label{eq:Schi}
\eea
with $\mK$, $\mU$, $\tilde{\alpha}$ and $\tilde{\beta}$ some functions of $\varphi$ that will be determined from our requirement and $W(\chi)$ is some potential for the scalar $\chi$. In our Ansatz we have included the self-interactions of the scalar field encoded in $\mK$ and $\mU$. Since this sector was resolved above, we will focus here on the couplings to $\chi$ so that $\tilde{\alpha}$ and $\tilde{\beta}$ are the functions to be determined by imposing the $\varphi$ field equations be of the form
\bea
\square\varphi+V'(\varphi)=-\frac{1}{\Msc}T.
\eea
The trace of the total energy-momentum tensor derived from (\ref{eq:Schi}) is given by
\bea
T=4\mU(\varphi) V(\varphi)+4\tilde{\beta}(\varphi) W(\chi)-\mK(\varphi) \partial_\mu\varphi\partial^\mu\varphi-\tilde{\alpha}(\varphi) \partial_\mu\chi\partial^\mu\chi
\eea
and hence we obtain that the equation of motion must be of the form
\bea
\square\varphi+V'(\varphi)=-\frac{1}{\Msc}(4\mU(\varphi) V(\varphi)+4\tilde{\beta}(\varphi) W(\chi)-\mK(\varphi) \partial_\mu\varphi\partial^\mu\varphi-\tilde{\alpha}(\varphi) \partial_\mu\chi\partial^\mu\chi).
\label{eq:phichi1}
\eea
On the other hand, varying (\ref{eq:Schi}) with respect to $\varphi$ yields
\bea
\square\varphi=-\frac{1}{\mK(\varphi)}\left(\frac12\mK'(\varphi)\partial_\mu\varphi\partial^\mu\varphi-(\mU(\varphi) V(\varphi))'+\frac12\tilde{\alpha}'(\varphi)\partial_\mu\chi\partial^\mu\chi-\tilde{\beta}'(\varphi)W(\chi)\right).
\label{eq:phichi2}
\eea
Comparing (\ref{eq:phichi1}) and (\ref{eq:phichi2}) will give the equations that must be satisfied by the functions in our Ansatz for the action. The $\varphi$ sector has already been solved in the previous subsection, so we will only pay attention to the $\chi$ sector now. Then, we see that the functions $\tilde{\alpha}$ and $\tilde{\beta}$ must satisfy the following equations
\be
\frac{\tilde{\alpha}'}{\tilde{\alpha}}=\frac{2\mK}{\Msc},\quad\quad\frac{\tilde{\beta}'}{\tilde{\beta}}=\frac{4\mK}{\Msc}.
\ee
The solution for these equations, taking into account the functional form of $\mK(\varphi)$ given in \ref{eq:solKsource}, is then
\be
\tilde{\alpha} = 1+\frac{2\varphi}{\Msc},\quad\quad\tilde{\beta} = \left(1+\frac{2\varphi}{\Msc}\right)^2
\ee
that coincides with the expression given in \cite{Freund:1969hh}. We see again that, although both procedures give the same leading order coupling to matter for the scalar field, the full theory crucially depends on whether the coupling is imposed at the level of the action or the equations. If we consider again the case of a cosmological constant as the matter sector, we see that its re-dressing with the scalar field will be different in both cases. It could be interesting to explore the differences with respect to the analysis performed in \cite{Sami:2002se}, where the coupling was assumed to occur at the level of the action.

\section{Vector gravity}
After having revisited and extended the case of a scalar field coupled to the energy-momentum tensor, we now turn to the case of a vector field. Since vectors present a richer structure than scalars due to the possibility of having a gauge invariance or not depending on whether the vector field is massless of massive, we will distinguish between gauge invariant couplings and non-gauge invariant couplings. For the latter, the existence of a decoupling limit where the dominant interactions correspond to those of the longitudinal mode will lead to a resemblance between some of the interactions obtained here and those of the derivatively coupled scalar studied above.

\subsection{Self-coupled Proca field}
Analogously to the scalar field case, our starting point will be the action for a massive vector field given by the Proca action\footnote{Of course we could consider an arbitrary potential, but a mass term is the natural choice if we really assume that we start with a free theory.}
\be
\mS_{(0)}=\int\d^4x\left(-\frac14F_{\mu\nu}F^{\mu\nu}+\frac12m^2A^2\right)
\ee
where $F_{\mu\nu}=\partial_\mu A_\nu-\partial_\nu A_\mu$, $A^2\equiv A_\mu A^\mu$ and $m^2$ is the mass of the vector field. The energy-momentum tensor of this field is given by
\be
T_{(0)}^{\mu\nu}\,=\,-F^{\mu\alpha}F^{\nu}{}^{\alpha}+\frac{1}{4}\eta^{\mu\nu}F_{\alpha\beta}F^{\alpha\beta}-\frac{m^2}{2}\eta^{\mu\nu}A^2+m^2 A^{\mu}A^{\nu}\,.
\ee
Unlike the case of the scalar field, this energy-momentum tensor does not coincide with the canonical one obtained from Noether's theorem and, thus, the Belinfante-Rosenfeld procedure would be needed to obtain a symmetric energy-momentum tensor, showing the importance of the choice in the definition of the energy-momentum tensor in the general case.

Along the lines of the procedure carried out in the previous sections, we will now introduce self-interactions of the vector field by coupling it to its energy-momentum tensor, so the first correction will be
\be
\mS_{(1)}=\frac{1}{\Mvc^2}\int\d^4xA_\mu A_\nu T_{(0)}^{\mu\nu}
\ee
with $\Mvc^2$ the corresponding coupling scale. In this case, the leading order interaction corresponds to a dimension 6 operator. Since this interaction will also contribute to the energy-momentum tensor, we will need to add yet another correction as in the previous cases, resulting in an infinite series in $A^2/\Mvc^2$ that reads:
\bea
\mS=\int\d^4x\Big[&-&\frac14\Big(1-Y-Y^2-3Y^3+\cdots \Big)F_{\mu\nu}F^{\mu\nu}+\frac12\Big(1+Y+3Y^2+15Y^3+\cdots   \Big)m^2A^2\nonumber\\
&-&\frac{1}{\Mvc^2}\Big(1+2Y+9Y^2+\cdots\Big)A_\mu A_\nu F^{\mu\alpha}F^\nu{}_\alpha\Big]
\eea
where $Y\equiv A^2/\Mvc^2$.
Again, to resum the iterative process we will use a guessed form for the full action. The above perturbative series makes clear that the final form of the action will take the form
\be
\mS=\int\d^4x\left[-\frac14\alpha(Y)F_{\mu\nu}F^{\mu\nu}-\frac{1}{\Mvc^2}\beta(Y)A_\mu A_\nu F^{\mu\alpha}F^\nu{}_\alpha+\frac{m^2}{2}\mU(Y)A^2\right]\nonumber\\
\ee
where the functions $\alpha$, $\beta$ and $\mU$ will be obtained by imposing the desired form of the interactions through the total energy-momentum tensor, i.e., we need to have
\bea
\mS&=&\int\d^4x\left[-\frac14\alpha(Y)F_{\mu\nu}F^{\mu\nu}-\frac{1}{\Mvc^2}\beta(Y)A_\mu A_\nu F^{\mu\alpha}F^\nu{}_\alpha+\frac{m^2}{2}\mU(Y)A^2\right]\nonumber\\
&=&\int\d^4x\left[-\frac14F_{\mu\nu}F^{\mu\nu}+\frac12m^2A^2+\frac{1}{\Mvc^2}A_\mu A_\nu T^{\mu\nu}\right]\nonumber\\
&=&\int\d^4x\left[-\frac14\Big(1-Y\alpha(Y)+2Y^2\alpha'(Y) \Big)F_{\mu\nu}F^{\mu\nu}+\frac{m^2}{2}\Big(1+Y\mU(Y)+2Y^2\mU'(Y)\Big)A^2\right.\nonumber\\
&&\quad\quad\quad\;\left.-\frac{1}{\Mvc^2}\Big(\alpha(Y)+3Y\beta(Y) +2Y^2\beta'(Y)  \Big)A_\mu A_\nu F^{\mu\alpha}F^\nu{}_\alpha\right].
\eea
Thus, the coupling functions in the resummed action need to satisfy the following first order differential equations
\bea
\alpha&=& 1-Y\alpha+2Y^2\alpha'\;,\\
\beta&=&\alpha+3Y\beta+2 Y^2\beta'\;,\\
\mU&=&1+Y\mU+2Y^2\mU'\; .
\eea
These equations are of the same form as the ones obtained for the derivatively coupled scalar field and the solutions will also present similar features. For instance, the perturbative series will need to be interpreted as asymptotic expansions of the solutions of the above equations. Furthermore, the integration constants that determine the desired solution are already implemented in the equations so we need to impose regularity at the origin, but this only selects a unique solution for a given semi-axis, either $Y>0$ or $Y<0$, that can then be matched to an infinite family of solutions in the complementary semi-axis (see Appendix \ref{Appendix}). The particular form of the solutions is not specially relevant for us here (although it will be relevant for practical applications), but we will only remark that they correspond to the class of theories for a vector field that is quadratic in the field strength or, in other words, in which the field strength only enters linearly in the equations.

As for the derivative couplings of the scalar field, we can be more general and allow for a coupling of the form
\be
\mS_{(1)}=\frac{1}{\Mvc^2}\int\d^4x\Big(b_1A_\mu A_\nu+b_2A^2\eta_{\mu\nu}\Big) T_{(0)}^{\mu\nu}.
\ee
The iterative process in this case gives rise to
\bea
\mS=\int\d^4x\Big[&-&\frac14\Big(1-b_1Y-b_1(b_1+2b_2)Y^2-b_1(b_1+2b_2)(3b_1+4b_2)Y^3+\cdots \Big)F_{\mu\nu}F^{\mu\nu}\nonumber\\
&-&\frac{b_1}{m^2}\Big(1+2(b_1+b_2)Y+(9b_1^2+16b_1b_2+8b_2^2)Y^2+\cdots\Big)A_\mu A_\nu F^{\mu\alpha}F^\nu{}_\alpha\nonumber\\
&+&\frac12\Big(1+(b_1-2b_2)Y+3b_1(b_1-2b_2)Y^2+3b_1(b_1-2b_2)(5b_1+2b_2)Y^3+\cdots   \Big)m^2A^2\Big].\nonumber
\eea
We can again use our Ansatz for the final action to obtain that the differential equations to be satisfied are
\bea
\alpha&=& 1-b_1Y\alpha+2(b_1+b_2)Y^2\alpha'\\
\beta&=&b_1\alpha+(3b_1+2b_2)Y\beta+2(b_1+b_2) Y^2\beta'\\
\mU&=&1+(b_1-2b_2)Y\mU+2(b_1+b_2)Y^2\mU'\; .
\eea
Remarkably, we see from the perturbative expansion that the case $b_1=0$ exactly cancels all the corrections to the kinetic part so that $\alpha=1$, $\beta=0$ and only the potential sector is modified. It is not difficult to check that this is indeed a solution to the above differential equations. We can also see here again that the choice $b_1+b_2=0$, which corresponds to a coupling to the orthogonal projector to the vector field given by $\eta_{\mu\nu}-A_\mu A_\nu/A^2$, reduces the equations to a set of algebraic equations whose solution is
\bea
\alpha&=&\frac{1}{1+b_1Y},\\
\beta&=&\frac{b_1\alpha}{1-b_1Y}=\frac{b_1}{1-b_1^2Y^2},\\
\mU&=&\frac{1}{1-3b_1Y}.
\eea
These solutions show that $\alpha$ and $\beta$ present different analytic properties depending on whether the field configuration is timelike or spacelike, but, in any case, $\beta$ has a pole for $\vert b_1 Y\vert =1$ so that it seems reasonable to demand $\vert b_1Y\vert<1$ for this particular solution. The properties of these equations are similar to the ones we found in Sec. \ref{Sec:ScalarDerivative} for the derivatively coupled case and, in fact, the equation for $\mU$ here is the same as the equation for $\mK$ in (\ref{Eq:eqsderivativecoupling}), which is of course no coincidence. Thus, the more detailed discussion given in Appendix \ref{Appendix} also applies and, in particular, it will also be possible to obtain polynomial solutions by appropriately choosing the parameters, owed to the recursive procedure used to construct the interactions.

From our general solution we can also analyse what happens if our starting free theory is simply a Maxwell field, i.e., $m^2=0$. In that case, only the terms containing $\alpha$ and $\beta$ will have an effect and, in fact, they will provide the vector field with a mass around non-trivial backgrounds of $F_{\mu\nu}$, signaling that the number of perturbative propagating polarisations will depend on the background configuration. However, the lack of a gauge symmetry in the full theory makes a vanishing bare mass seem like an unnatural choice.

So far we have solved the problem of a vector field coupled to its own energy-momentum tensor at the full non-linear level in the action. We could also follow the procedure of finding the theory such that the energy momentum-tensor is the source of the vector field equations of motion. This is analogous to the case of scalar gravity considered above and also the procedure that leads to GR for the spin-2 case. However, a crucial difference arises for the vector field\footnote{The discussion presented here also applies to the case of derivative couplings for the scalar field.} case owed to the fact that the leading order interaction is quadratic in the vector field and, thus, the energy-momentum tensor cannot act as a source of the vector field equations. We have different possibilities then on how to generalise the linear coupling to the energy-momentum tensor to the full theory at the level of the field equations. Because of the lack of a clear criterion at this point and the existence of several different inequivalent possibilities of carrying out this procedure, we will not pursue it further here and we will content ourselves with the analysis of the construction of the theories where the coupling occurs at the level of the action. We will simply mention that a possibility to get this difficulty around is by breaking Lorentz invariance. If we introduce some fixed vector $u^\mu$ (that could be identified for instance with some vev of the vector field), then we could construct our interactions as e.g. $A_\mu T^{\mu\nu}u_\nu$ so that $T^{\mu\nu}u_\nu$ would act as the source of the vector field equations and, then, we could extend this result at the non-linear level.

\subsection{Derivative gauge-invariant self-couplings}

In the previous section we have studied the case where the vector field couples to its energy-momentum tensor without imposing gauge invariance. For a scalar field, the equations of motion do not contain any off-shell conserved current  derived from some Bianchi identities, and this makes a crucial difference with respect to the vector field case where we do have the Bianchi identities derived from the $U(1)$ symmetry of the Maxwell Lagrangian. Even if we break the gauge symmetry by adding a mass term, the off-shell current leads to a constraint equation that must be satisfied. In the theories obtained in the previous section by coupling $A_\mu$ to the energy-momentum tensor, we also obtain a constraint equation by taking the divergence of the corresponding field equations. This constraint is actually the responsible for keeping three propagating degrees of freedom in the theory. In fact, starting from a massless vector field with its gauge invariance, the couplings actually generate a mass term around non-trivial backgrounds, increasing that way the number of polarisations. In this section, we will aim to re-consider our construction by maintaining the $U(1)$ gauge symmetry of Maxwell theory also in the couplings of the vector field to its own energy-momentum tensor. This also resembles somewhat the extension of the scalar field case to include derivative couplings arising from imposing a shift symmetry, although with crucial differences, for instance the scalar field interactions are based on a global symmetry while the ones considered in this section will be dictated by a gauge symmetry. 

After the above clarifications on the procedure that we will follow in this section, we can write our starting action describing a massless vector field
\be
\mS_{(0)}=-\frac14\int\d^4x F_{\mu\nu} F^{\mu\nu},
\ee
and add interactions to its own energy-momentum tensor respecting the $U(1)$ symmetry. At the lowest order, these interactions can be written as
\be
\mS_{(1)}=\frac{1}{\MF^4}\int\d^4x \Big(b_1F_{\mu\alpha} F_\nu{}^\alpha +b_2 F_{\alpha\beta} F^{\alpha\beta}\eta_{\mu\nu}\Big)T^{\mu\nu}_{(0)}
\ee
with $\MF$ the corresponding coupling scale, $b_{1,2}$ dimensionless parameters and the Maxwell energy-momentum tensor given by
\be
T_{(0)}^{\mu\nu}\,=\,-F^{\mu\alpha}F^{\nu}{}_{\alpha}+\frac{1}{4}\eta^{\mu\nu}F_{\alpha\beta}F^{\alpha\beta}.
\ee
Before proceeding further, let us digress on the form of the introduced interactions. We have followed the easiest possible path of adding couplings that trivially respect the original $U(1)$ symmetry by using the already {\it gauge invariant} field strength $F_{\mu\nu}$. This is along the lines of the Pauli interaction term for a charged fermion $\psi$ given by $\bar{\psi}[\gamma^\mu,\gamma^\nu]\psi F_{\mu\nu}$, where gauge invariance is trivially realised. Of course, this non-renormalisable term is within the effective field theory, but the usual renormalisable coupling $A_\mu J^\mu$ gives the leading order interaction. This is somehow analogous to existing constructions for the spin-2 case where one seeks for a consistent coupling of the graviton to the energy-momentum tensor respecting the original linearised diffeomorphisms invariance. Besides the usual coupling $h_{\mu\nu}T^{\mu\nu}$ that respects the symmetry upon conservation of $T^{\mu\nu}$, the realisation of the linearised diffeomorphisms can be achieved in two different ways, namely: one can either introduce a coupling of the form $h_{\mu\nu} P^{\mu\nu\alpha\beta} T_{\alpha\beta}$ with $P^{\mu\nu\alpha\beta}$ an identically divergenceless projector $\partial_\mu P^{\mu\nu\alpha\beta}=0$ (see for instance \cite{Deser:1968zza}), or one can add a coupling of the matter fields to an exactly gauge invariant quantity. The second approach is possible by using that the linearised Riemann tensor $R^{\rm L}_{\mu\nu\alpha\beta}(h)$ is {\it gauge invariant}, and not only covariant (see the seminal Wald's paper \cite{Wald:1986bj} and 
\cite{Hertzberg:2016djj} for interesting discussions on these alternative couplings). This is the analogous construction scheme we are following here. One could have also tried to follow another approach and tried to construct non-derivative couplings with a field dependent realisation of the original $U(1)$ symmetry whose lowest order in fields is given by the usual transformation law. The non-linear completion of gauge symmetry can correspond either to the original $U(1)$ symmetry up to some field redefinition or to a genuine non-linear completion, what will likely require transformations involving higher derivatives of the gauge parameter (see for instance \cite{Wald:1986bj}). This interesting path will not be pursued further here and we will focus on the simplest case.

After briefly discussing some alternatives for gauge invariant couplings, we will proceed with the construction considered here. The iterative process in this case leads to the perturbative series
\bea
\mS&=&-\frac14\int\d^4x\left[\Big(1+b_1 Z+b_1(3b_1+4b_2)Z^2+b_1(3b_1+4b_2)^2 Z^3 +\cdots\Big)F_{\mu\nu}F^{\mu\nu}\right.\\
&&+\left.\frac{b_1}{\MF^4}\Big(1+(3b_1+4b_2)Z+(3b_1+4b_2)(5b_1+8b_2)Z^2+b_1(3b_1+4b_2)\tilde{Z}^2+\cdots\Big)\big(F_{\mu\nu} \tilde{F}^{\mu\nu} \big)^2\right]\nonumber
\eea
where $Z\equiv F_{\alpha\beta}F^{\alpha\beta}/\MF^4$, $\tilde{Z}\equiv F_{\alpha\beta}\tilde{F}^{\alpha\beta}/\MF^4$ and $\tilde{F}^{\mu\nu}=\frac12\epsilon^{\mu\nu\alpha\beta}F_{\alpha\beta}$ is the dual of the field strength. Moreover, we have used the identity $4F^\mu{}_\nu F^\nu{}_\rho F^\rho{}_\sigma F^\sigma{}_\mu=2(F_{\mu\nu}F^{\mu\nu})^2+(F_{\mu\nu}\tilde{F}^{\mu\nu})^2$. 
This perturbative series is not straightforwardly resummed, so we will follow the alternative procedure of making an Ansatz for the resummed action. Since we are maintaining gauge invariance the resulting action must be a function of the two independent Lorentz and gauge invariants, i.e., the action must take the form
\be
\mS=M_F^4\int \d^4x \,\mK(Z,\tilde{Z}),
\ee
with $\mK$ a function to be determined. Notice that our prescribed interactions do not break parity so that $\mK$ will need to be an even function of $\tilde{Z}$. Given our requirement of a coupling to the energy-momentum tensor, the action also needs to take the form
\begin{align}
\mS & =\int \d^4x\[-\frac{1}{4}F_{\mu\nu}F^{\mu\nu}+\frac{1}{M_F^4}\(b_1 F_{\mu\alpha}F_\nu{}^\alpha+b_2F_{\alpha\beta}F^{\alpha\beta}\eta_{\mu\nu}\)T^{\mu\nu}\]\notag\\
& = M_F^4\int\d^4 x\[-\frac{Z}{4}-(b_1+4b_2)Z(\mK-\tilde{Z}\mK_{\tilde{Z}})+(2(b_1+2b_2)Z^2+b_1\tilde{Z}^2)\mK_Z\].
\end{align}
From these two expressions we conclude that $\mK(Z,\tilde{Z})$ must satisfy the partial differential equation:
\be
\mK=-\frac{Z}{4}-(b_1+4b_2)Z\Big(\mK-\tilde{Z}\mK_{\tilde{Z}}\Big)+\Big[2(b_1+2b_2)Z^2+b_1\tilde{Z}^2\Big]\mK_Z.
\label{Eq:ZZt}
\ee
We can easily check that for $b_1=0$ the perturbative series does not generate any interaction, which is nothing but a reflection of the conformal invariance of the Maxwell Lagrangian in 4 dimensions that leads to a traceless energy-momentum tensor. If we consider that case, the above equation reduces to $\mK=-Z/4+4b_2(-\mK+Z\mK_Z+\tilde{Z}\mK_{\tilde{Z}})$, that has the Maxwell Lagrangian as a regular solution. On the other hand, for $3b_1+4b_2=0$, we see that only the first corrections in the perturbative expansion remains so that one expects the full solution to take the simple polynomial form $\mK=-1/4 \big[(1+b_1 Z) Z + b_1 \tilde{Z}^2\big]$. One can readily check that this is indeed the solution of Eq. (\ref{Eq:ZZt}) with $b_2=-3b_1/(4b_4)$. As in the previous sections dealing with derivatively coupled scalars or non-gauge invariant couplings for vectors, the existence of this polynomial solution is a direct consequence of the recursive procedure generating the interactions and, thus, we could also choose the parameters $b_1$ and $b_2$ as to have some higher order polynomial solutions for the gauge invariant theories obtained here.

\subsection{First order formalism}

In the previous section we have looked at the theory for a vector field that is coupled to its own energy-momentum tensor in a gauge invariant way. We have just seen that the gauge-invariant coupling leads to a partial differential equation that is, in general, not easy to solve so we will now consider the problem from the first order formalism perspective, hoping that it will simplify the resulting equations, as it happens in other contexts. Unfortunately, we will see that this does not seem to be the case here. The starting free theory for a $U(1)$ invariant vector field can be described in the first order formalism by the action
\bea
\mS_{(0)}=\int\d^4x\left[\Pi^{\mu\nu}\partial_{[\mu} A_{\nu]}+\frac14 \Pi^{\mu\nu}\Pi_{\mu\nu}\right].
\eea
Upon variations with respect to the momentum $\Pi^{\mu\nu}$ and the vector field, we obtain the usual De Donder-Weyl-Hamilton equations  $\Pi_{\mu\nu}=\partial_\nu A_\mu-\partial_\mu A_\nu=-F_{\mu\nu}$ and $\partial_\mu\Pi^{\mu\nu}=0$ respectively, that reproduce the usual Maxwell equations $\partial_\mu F^{\mu\nu}=0$. By integrating out the momentum, we reproduce the usual Maxwell theory in the second order formalism. At the lowest order we then prescribe the self-interaction be of the form
\bea
\mS_{(0)}+\mS_{(1)}=\int\d^4x\left[\Pi^{\mu\nu}\partial_{[\mu}A_{\nu]}+\frac14 \Pi^{\mu\nu}\Pi_{\mu\nu}+\frac{1}{\MF^4}\big(b_1 \Pi_{\mu\alpha}\Pi_\nu{}^\alpha+b_2\Pi^{\alpha\beta}\Pi_{\alpha\beta}\,\eta_{\mu\nu}\big)T_{(0)}^{\mu\nu}\right]
\eea
with $T_{(0)}^{\mu\nu}$ the energy-momentum tensor corresponding to the free theory $\mS_{(0)}$. This is the form that we will require for the final theory replacing $T_{(0)}^{\mu\nu}$ by the total energy-momentum tensor. Following the same reasoning as in the previous section, the final theory should be described by the following action
\bea
\mS=\int\d^4x\left[\Pi^{\mu\nu}\partial_{[\mu}A_{\nu]}-\Ham(Z,\tilde{Z})\right]
\label{Eq:SAnsatzfirstorder}
\eea
where the Hamiltonian $\Ham$ must be a function of the momentum due to gauge invariance and, furthermore, Lorentz invariance imposes that the dependence must be through the only two independent Lorentz scalars, namely $Z=\Pi^{\mu\nu}\Pi_{\mu\nu}/\MF^4$ and $\tilde{Z}=\Pi^{\mu\nu}\tilde{\Pi}_{\mu\nu}/\MF^4$. Parity will also impose it to be an even function of $\tilde{Z}$. In order to proceed to compute the energy-momentum tensor, we need to choose the tensorial character of the phase space variables and, as we discussed in Sec. \ref{Sec:ScalarFirstOrder}, this may lead to substantial simplifications. As in Sec. \ref{Sec:ScalarFirstOrder}, let us assume that $\Pi^{\mu\nu}$ is a density so that $P^{\mu\nu}=\Pi^{\mu\nu}/\sqrt{-\gamma}$ is a zero-weight tensor. This has the advantage that the term $\Pi^{\mu\nu}\partial_{[\mu}A_{\nu]}$ in (\ref{Eq:SAnsatzfirstorder}) will not contribute to $T^{\mu\nu}$. The total energy-momentum tensor is thus given by
\be
T^{\mu\nu}=\left[\Ham-2\frac{\partial \Ham}{\partial Z}\Pi^2-\frac{\partial \Ham}{\partial \tilde{Z}}\Pi^{\alpha\beta}\tilde{\Pi}_{\alpha\beta}\right]\eta^{\mu\nu}+4\frac{\partial \Ham}{\partial Z}\Pi^{\mu\alpha}\Pi^\nu{}_\alpha
\label{Eq:Tvectorfirstorder}
\ee
so that the interaction term reads
\be
\frac{1}{\MF^4}\big(b_1 \Pi_{\mu\alpha}\Pi_\nu{}^\alpha+b_2\Pi^{\alpha\beta}\Pi_{\alpha\beta}\,\eta_{\mu\nu}\big)T^{\mu\nu}=(b_1+4b_2)Z\left(\Ham-\tilde{Z}\frac{\partial \Ham}{\partial \tilde{Z}}\right)+\left[b_1\tilde{Z}^2-4b_2Z^2\right]\frac{\partial \Ham}{\partial Z}.
\ee
From this expression it is already apparent that the resummation will necessarily involve the resolution of a partial differential equation as in the first order formalism case. In fact, the resulting equation will be of the same type and, consequently, resorting to the first order formalism does not lead to any simplification. One may think that another choice of the weight for the momentum could lead to some simplifications, but that is not the case and, in fact, choosing an arbitrary weight leads to the same result. To show this more explicitly, let us assume that the momentum has an arbitrary weight $w$ so that $P^{\mu\nu}=(\sqrt{-\gamma})^{-w}\Pi^{\mu\nu}$ is a tensor of zero weight. Then, the variation of the Hamiltonian with respect to the auxiliary metric will give
\be
\delta\Ham=-\frac{\partial\Ham}{\partial Z}\Big(w\Pi^2\gamma^{\mu\nu}-2\Pi^{\mu\alpha}\Pi^\nu{}_\alpha\Big)\delta\gamma_{\mu\nu}+\frac12\frac{\partial\Ham}{\partial \tilde{Z}}(1-2w)\Pi^{\alpha\beta}\tilde{\Pi}_{\alpha\beta}\gamma^{\mu\nu}\delta\gamma_{\mu\nu}
\ee
where we have taken into account that the Hamiltonian $\Ham$ becomes a function $\Ham\big(Z,\tilde{Z})\to\Ham\big(\vert\gamma\vert^{-w/2}Z,\vert\gamma\vert^{-w/2}\tilde{Z}\big)$ after our covariantisation choice. The total energy-momentum tensor can be readily computed to be
\be
T^{\mu\nu}=\left[(w-1)\Pi^{\alpha\beta}\partial_{[\alpha}A_{\beta]}+\Ham-2w\frac{\partial\Ham}{\partial Z}\Pi^2+(1-2w)\frac{\partial\Ham}{\partial \tilde{Z}}\Pi^{\alpha\beta}\tilde{\Pi}_{\alpha\beta}\right]\eta^{\mu\nu}+4\frac{\partial\Ham}{\partial Z}\Pi^{\mu\alpha}\Pi^\nu{}_\alpha\, .
\label{Eq:Tgeneralw}
\ee
It is easy to see that this expression directly gives the energy-momentum tensor of a Maxwell field with $\Ham=-\frac14\Pi^2$ if we set $w=0$ and integrate out the momentum by using $\Pi^{\mu\nu}=-F^{\mu\nu}$. For the general case, we need to express the energy-momentum tensor in phase space variables. From the equation for $\Pi^{\mu\nu}$ we have
\be
\partial_{[\mu}A_{\nu]}=\frac{\partial \Ham}{\partial \Pi^{\mu\nu}}=2\frac{\partial\Ham}{\partial Z}\Pi_{\mu\nu}+\frac{\partial\Ham}{\partial \tilde{Z}}\tilde{\Pi}_{\mu\nu}.
\ee
If we insert this expression into (\ref{Eq:Tgeneralw}) we obtain
\be
T^{\mu\nu}=\left[\Ham-2\frac{\partial \Ham}{\partial Z}\Pi^2-\frac{\partial \Ham}{\partial \tilde{Z}}\Pi^{\alpha\beta}\tilde{\Pi}_{\alpha\beta}\right]\eta^{\mu\nu}+4\frac{\partial \Ham}{\partial Z}\Pi^{\mu\alpha}\Pi^\nu{}_\alpha,
\ee
which does not depend on the weight and, therefore, it is exactly the same that we obtained in (\ref{Eq:Tvectorfirstorder}). Of course, this is not very surprising, since, as a consequence of the argument in footnote \ref{F7}, different choices of weights only result in quantities that vanish on-shell. In this case, the use of the equation of motion of $\Pi^{\mu\nu}$ in order to express $T^{\mu\nu}$ in phase space variables precisely corresponds to the mentioned on-shell difference. 

For completeness, let us give the resulting equation in this case
\be
\Ham=-\frac{\MF^4}{4}Z-(b_1+4b_2)Z\left(\Ham-\tilde{Z}\frac{\partial \Ham}{\partial \tilde{Z}}\right)-\left[b_1\tilde{Z}^2-4b_2Z^2\right]\frac{\partial \Ham}{\partial Z}.
\ee
We see that, as advertised, the first order formalism does not seem to give any advantage with respect to the second order formalism in this case. It may be that there is some clever choice of phase space coordinates that does reduce the difficulty of the problem.

\subsection{Coupling to matter fields}\label{Sec:VectorMatter}

We will end our study of the vector field case by considering couplings to matter fields through the energy-momentum tensor. As in Sec. \ref{Sec:Scalarmattercoupling} we will take a scalar field $\chi$ as a proxy for the matter. Moreover, as we mentioned in $\ref{Sec:Scalarmattercoupling}$, the results obtained here will also give how the scalar field $\varphi$ couples to matter fields when the interactions follow our prescription. Thus, our starting action will now contain the additional term
\be
\mS_{\chi,\rm (1)}=\int\d^4x\left[\frac12\partial_\mu\chi\partial^\mu\chi-W(\chi)+\frac{1}{\Mvd^2}\Big(b_1A_\mu A_\nu+b_2A^2\eta_{\mu\nu}\Big) T^{\mu\nu}\right]
\ee
where $T^{\mu\nu}$ includes the energy-momentum of the own vector field plus the contribution coming from the scalar field. The iterative process applied to this case yields the following expansion
\begin{align}
\mS_\chi=&\int\d^4x\left[\frac12\mathcal{K}^{\mu\nu}\partial_\mu\chi\partial_\nu\chi -\left(1-(b_1+4b_2)\left(Y+(b_1-2b_2)Y^2+3b_1(b_1-2b_2)Y^3 \cdots\right) \right)W(\chi) \right]
\end{align}
where we have defined
\bea
\mathcal{K}^{\mu\nu}\equiv&&\left[1-(b_1+2b_2)\left(Y+b_1Y^2+b_1(3b_1+2b_2)Y^3 \cdots \right)\right]\eta^{\mu\nu}\nonumber\\
&&+2b_1\left(1+2(b_1-b_2)Y+(9b_1^2-8b_1b_2-4b_2^2)Y^2 \cdots\right)A^{\mu}A^\nu.
\eea
In this case, our Ansatz for the resummed action is
\be
\mS_\chi=\int\d^4x\left[\frac12\Big(C(Y)\eta^{\mu\nu}+D(Y)A^\mu A^\nu\Big)\partial_\mu\chi\partial_\nu\chi-U(Y)W(\chi)\right]
\ee
so we have that the action reads
\begin{align}
\mS_\chi&=\int\d^4x\left[\frac12\partial_\mu\chi\partial^\mu\chi-V(\chi)+\frac{1}{\Mvd^2}(b_1A_\mu A_\nu+b_2 A^2\eta_{\mu\nu}) T^{\mu\nu}\right]\nonumber\\
&=\int\d^4x\left[\frac12\left(1-(b_1+2b_2)YC+2(b_1+b_2)Y^2C'\right)\partial_\mu\chi\partial^\mu\chi\right.\nonumber\\
&\hspace{2cm}+\frac12\left(2b_1C+3b_1YD+2(b_1+b_2)Y^2D'  \right)A^\mu A^\nu\partial_\mu\chi\partial_\nu\chi\nonumber\\
&\hspace{2cm}\left.-\left(1+(b_1+4b_2)YU-2(b_1+b_2)Y^2U'  \right)W(\chi)\right]
\end{align}
and, therefore, the functions $C$, $D$ and $U$ should be the solutions of
\bea
C&=&1-(b_1+2b_2)YC+2(b_1+b_2)Y^2C',\\
D&=&2b_1C+3b_1YD+2(b_1+b_2)Y^2D' ,\\
U&=&1+(b_1+4b_2)YU-2(b_1+b_2)Y^2U' .
\eea
We see again that when coupling to the orthogonal projector, i.e., $b_2=-b_1$, the equations become algebraic and the solution is
\bea
C&=&\frac{1}{1-b_1Y},\\
D&=&\frac{1+2b_1(1-Y)}{1-4b_1Y+3b_1^2Y^2},\\
U&=&\frac{1}{1+3b_1Y}.
\eea
The form of the equations are similar to the ones found in the precedent sections so we will not repeat once again the same discussion, but obviously the same types of solutions will exist in this case. Let us however mention that the same results for the matter coupling can be obtained for the derivatively coupled scalar field upon the replacement $A_\mu\to\partial_\mu\varphi$.

\section{Superpotential terms}
\label{sec:Superpotentials}

In the previous sections we have considered the energy-momentum tensor obtained from the usual prescription of coupling it to gravity and taking variational derivatives with respect to the metric. However, as we already explained above, the energy-momentum tensor (as any usual Noether current) admits the addition of super-potential terms with vanishing divergence either identically or on-shell. This freedom in the definition of the energy-momentum tensor can be used to {\it improve} the canonical energy-momentum tensor in special cases. For instance, theories involving spin 1 fields give non-symmetric canonical energy-momentum tensors that can be symmetrised by adding suitable superpotentials given in terms of the generators of the corresponding Lorentz representation (the Belinfante-Rosenfeld procedure). If the theory has a gauge symmetry, one can also add super-potential terms (on-shell divergenceless this time) to obtain a gauge invariant energy-momentum tensor\footnote{Let us recall the Weinberg-Witten theorem here that prevents the construction of a Lorentz covariant and gauge invariant energy-momentum tensor for particles with spin $\geq 2$. } obtaining then the energy-momentum tensor that results from the Rosenfeld prescription. Theories featuring scale invariance admit yet another improvement to make the energy-momentum tensor traceless\footnote{With only scale invariance the trace of the energy-momentum tensor is given by the divergence of a vector and only when the theory exhibits full conformal invariance the energy-momentum tensor can be made traceless. Since theories that are scale invariant are also conformally invariant, we do not make a distinction here.}. This traceless energy-momentum tensor is not the one obtained from the Hilbert prescription (nor with the Belinfante-Rosenfeld procedure) upon minimal coupling to gravity, but one needs to add a non-minimal coupling to the curvature, which simply tells us that the iterative coupling procedure to gravity starting from the improved energy-momentum tensor gives rise to non-minimal couplings. This example illustrates how considering different super-potential terms can result in different theories for the full action. In this section we will briefly discuss this point within our constructions for the self-interactions of scalar and vector fields to their own energy-momentum tensors.

Let us start with the scalar field and consider a particular family of terms that lead to interesting results. Thes lowest order object that is identically divergence-free is given by
\be
\frac{1}{\Msd}X_{1}^{\mu\nu}=\Box\varphi \eta^{\mu\nu}-\partial^\mu\partial^\nu\varphi,
\ee
where we have introduced the factor $\Msd$ to match the dimension of an energy-momentum tensor for $X_1^{\mu\nu}$. This corresponds to a super-potential term of the form $\partial_\alpha(\partial^{[\alpha}\eta^{\mu]\nu})$. Now we want to study the effect on the full theory of adding this boundary term to the energy-momentum tensor. Notice that this object will give rise to an operator of lower dimensionality than the coupling to the energy-momentum tensor and, therefore, the added correction will be suppressed by one less power of the corresponding scale. In the case of the conformal coupling, we can see that the interaction $\varphi X^\mu{}_\mu$ simply amounts to a re-scaling of the kinetic term, so we will move directly to the derivative coupling. In that case, the first order correction is given by
\be
\Lag_{(1)}=\frac{1}{\Msd^4}\partial_\mu\varphi\partial_\nu\varphi X_{1}^{\mu\nu}=\frac{1}{\Msd^3}\Big[(\partial\varphi)^2\Box\varphi-\partial_\mu\varphi\partial_\nu\varphi \partial^\mu\partial^\nu\varphi\Big]=\frac{3}{2\Msd^3}(\partial\varphi)^2\Box\varphi
\ee
where we have integrated by parts and dropped a total derivative in the last term. We can recognize here the cubic Galileon Lagrangian \cite{Nicolis:2008in} that we have obtained by simply following our coupling prescription to the identically conserved object $X_{1}^{\mu\nu}$ that one can legitimately add to the energy-momentum tensor. By iterating the process we obtain the following perturbative expansion for the Lagrangian:
\be
\Lag=\big(1+3X+15X^2+105X^3+\cdots\big)\frac{1}{\Msd^3}\Big[(\partial\varphi)^2\Box\varphi-\partial_\mu\varphi\partial_\nu\varphi \partial^\mu\partial^\nu\varphi\Big].
\ee
We can now use that, for an arbitrary function $\mG(X)$ and upon integration by parts, we have 
\be
\mG(X)\partial_\mu\varphi\partial_\nu\varphi \partial^\mu\partial^\nu\varphi=\frac12\Msd^4\partial^\mu\varphi\partial_\mu \tilde{\mG}(X)\rightarrow -\frac12\tilde{\mG}(X)\Box\phi
\ee
with $\Msd^4\tilde{\mG}'(X)=\mG(X)$, to express the final Lagrangian as
\be
\Lag=\frac32\left(1+\frac52X+\frac{35}{3}X^2+\frac{315}{4}X^3+\cdots\right)\frac{(\partial\varphi)^2}{\Msd^3}\Box\varphi.
\ee
This Lagrangian is a particular case of the so-called KGB models \cite{Deffayet:2010qz} with a shift symmetry. Remarkably, we have obtained this Lagrangian from our construction, which can be understood in a similar fashion to the generation of non-minimal couplings in the case of gravity.

We can also obtain a {\it resummed} action by adding the superpotential term to the full theory so that its Lagrangian should read
\be
\Lag=\Lag_0+\frac{1}{\Msd^4}\Big[b_1\partial_\mu\varphi\partial_\nu\varphi+b_2 (\partial\varphi)^2\eta_{\mu\nu}\Big]\Big(T^{\mu\nu}+X^{\mu\nu}_1\Big)
\ee
being $T^{\mu\nu}$ the total energy-momentum tensor, i.e., the one computed from $\Lag$ by means of the Hilbert prescription. For the sake of generality, we have added here the more general coupling involving $b_1$ and $b_2$ as discussed in the precedent sections. In the above Lagrangian we have also included the term $\Lag_0$ that corresponds to the {\it free} Lagrangian, i.e., the part that survives in the decoupling limit $\Msd\rightarrow\infty$. This term will not affect the resummation of the interactions generated from the term with $X^{\mu\nu}_1$ so we do not need to consider it here (we already solved that part in the previous sections). Now we can notice that the interactions generated from the superpotential terms are of the form $\Lag=\Msd G(X)\Box\varphi$, up to boundary terms, so we need to have 
\be
\Lag_X=\Msd G(X)\Box\varphi=\Big[b_1\partial_\mu\varphi\partial_\nu\varphi+b_2 (\partial\varphi)^2\eta_{\mu\nu}\Big] \Big(X_{1}^{\mu\nu}+T_X^{\mu\nu}\Big)
\ee
being $T_X^{\mu\nu}$ the total energy-momentum tensor of $\Lag_X$. By computing this energy-momentum tensor we finally get
\be
\Lag_X=\Msd G(X)\Box\varphi=\Msd\left[\frac32X+2(b_1+b_2)X^2G'(X)+(b_1-2b_2)\tilde{G}(X)\right]\Box\varphi
\ee
with $\tilde{G}'(X)=XG'(X)$ and we have integrated by parts. From this relation we then see that the following equation must hold
\be
G(X)=\frac32X+2(b_1+b_2)X^2G'(X)+(b_1-2b_2)\tilde{G}(X).
\ee
Notice that this is an integro-differential equation, but we can easily transform it into the ordinary differential equation
\be
2(b_1+b_2)XF'(X)+\left(3b_1-\frac1X\right)F(X)+\frac32=0
\ee
with $F(X)\equiv XG'(X)$. As usual, the coupling to the orthogonal projector, i.e., $b_1=-b_2$ reduces the equations to a set of algebraic equations, although in this case an integration to go from $F$ to $G$ is necessary. In that case we obtain $F=-\frac32(3b_1-1/X)^{-1}$ so that $G=-1/(2b_1)\log(1-3b_1 X)$. In this case, the integration constant is irrelevant because it corresponds to a total derivative in the Lagrangian.

We have considered the simplest of the identically divergence-free superpotential terms, but we could consider the whole family given (in arbitrary dimension $d$) by
\be
X_{n}^{\mu\nu}\propto\Msd^{d-3n}\epsilon^{\mu\mu_1\cdots\mu_n\mu_{n+1}\cdots\mu_{d-1}}\epsilon^{\nu\nu_1\cdots\nu_n}{}_{\mu_{n+1}\cdots\mu_{d-1}}\partial_{\mu_1}\partial_{\nu_1}\varphi \cdots \partial_{\mu_n}\partial_{\nu_n}\varphi
\ee
which can be seen to be trivially divergence-free by virtue of the antisymmetry of the Levi-Civita tensor. These higher order superpotentials would generate higher order versions of the Galileon Lagrangians. At the first order, each $X_n^{\mu\nu}$ will obviously generate the $n$-th Galileon Lagrangian (in fact, Galileon fields are precisely defined by coupling the gradients of the scalar field to identically divergen-free objects), while the higher order corrections will eventually produce sub-classes of shift-symmetric generalised Galileons fields.

Similar results to those found for the scalar field can be obtained by adding superpotential terms in the case of vector fields. We can consider the lowest order and identically conserved object
\be
\frac{1}{\Mvc^2} Y_1^{\mu\nu}=\divA \eta^{\mu\nu}-\partial^\nu A^\mu
\ee
which corresponds to a superpotential of the form $\Theta^{\alpha\mu\nu}=2A^{[\alpha}\eta^{\mu]\nu}$. Unlike in the scalar field case, this superpotential is not symmetric in the last two indices. While in the scalar field case, a non-derivative coupling did not produce new terms, for the vector field case already algebraic couplings generate new interesting interactions (as expected because the non-derivative coupling for the vector is related to the derivative coupling for the scalar). The leading order interaction by coupling the vector to this superpotential term gives
\be
\Lag_{(0)}=A_\mu A_\nu Y^{\mu\nu}_1=\frac32 A^2\divA
\ee
where we have integrated by parts and dropped a total derivative term. We see that, as expected, we recover the cubic vector Galileon Lagrangian \cite{VectorGalileon}. It is remarkable that this interaction corresponds to a dimension four operator which means that it is not suppressed by the scale $\Mvc$ or, in other words, it will survive in the decoupling limit $\Mvc\rightarrow\infty$. It is not surprising the resemblance of this interaction with the case of the derivatively coupled scalar previously discussed and it is not difficult to convince oneself that the resummation will lead to the same type of equations. In addition, very much like in the scalar field case, there are higher order superpotential terms that can be constructed for the vector field case, but, given the similarities with the scalar field couplings, we will not give more details here. A more interesting class of super-potential terms would be those respecting the $U(1)$ gauge invariance. However, it is known that there are no Galileon-like interactions for abelian vector fields \cite{Deffayet:2013tca} so that we do not expect to find anything crucially new by adding gauge invariant superpotential terms, but similar interactions to the ones already worked out.

\section{Effective metrics and generating functionals}
\label{Sec:EffectiveMetrics}

In the precedent sections we have studied the coupling of scalar and vector fields to the energy-momentum tensor and how this can be generalised to the full theory. The definition that we have mostly considered for the energy-momentum tensor is the Hilbert prescription, although we have briefly commented on some interesting consequences of considering superpotential terms in the previous section. Since the Hilbert prescription gives the energy-momentum tensor as a functional derivative with respect to some fiducial metric, a natural question is to what extent the full theory can be expressed in terms of an effective metric. In this section we intend to briefly discuss this aspect with special emphasis in the cases considered throughout this work.

For the clarity of our construction, let us go back to the beginning and consider again the case of a conformally coupled scalar field considered in \ref{sec:Conformalscalar}. The starting point there was a scalar field coupled to the trace of the energy-momentum tensor as\footnote{In order to alleviate the notation in this section we will drop all the scales used in the precedent sections.}
\be
\Lag_{\rm int}^{(1)}=\varphi T_{(0)}.
\ee
It is not difficult to see that this interaction can be conveniently written as the following functional derivative
\be
\Lag_{\rm int}^{(1)}=\frac{\delta \mS_{(0)}[h_{\mu\nu}(J)]}{\delta J}\Big\vert_{J=0}
\ee
where the zeroth order action $\mS_{(0)}$ is evaluated on the conformal effective metric $h_{\mu\nu}=\exp(-2J\varphi)\eta_{\mu\nu}$ with $J$ an external field\footnote{We have chosen here the exponential for the conformal factor for simplicity, but any conformal factor $\Omega(J)$ satisfying $\Omega(0)=1$ and $\Omega'(0)=-2\varphi$ would do the job.}. The equivalence of the two expressions can be easily checked by using the chain rule:
\be
\frac{\delta \mS_{(0)}[h_{\mu\nu}(J)]}{\delta J}\Big\vert_{J=0}=\left(\frac{\delta \mS_{(0)}[h_{\mu\nu}(J)]}{\delta h_{\mu\nu}} \frac{\delta h_{\mu\nu}}{\delta J}\right)_{J=0}.
\ee
It is now immediate to re-obtain the results of Sec. \ref{sec:Conformalscalar} as well as obtaining generalisations. Let us assume that the initial action is a linear combination of homogeneous functions of the metric so we have
\be
\mS_{(0)}[\lambda\eta_{\mu\nu}]=\sum_i\lambda^{w_i} \mS_{(0),i}[\eta_{\mu\nu}],
\ee
with $\lambda$ some parameter and $w_i$ the degree of homogeneity of the corresponding term $\mS_{(0),i}$. Then, the functional derivative can be straightforwardly computed as
\be
\frac{\delta \mS_{(0)}[h_{\mu\nu}(J)]}{\delta J}\Big\vert_{J=0}=\sum_i\frac{\delta} {\delta J}\int\d^4x\exp(-2w_i\varphi J)\Lag_{(0),i}\Big\vert_{J=0}=\sum_i(-2w_i\varphi)\Lag_{(0),i}.
\ee
Since all the dependence on the metric in this first order correction is again in $\Lag_{(0),i}$ we have, for this specific case, that the $n$-th order interaction will be given by
\be
\Lag_{\rm int}^{(n)}=\frac{\delta \mS_{(n-1)}[h_{\mu\nu}(J)]}{\delta J}\Big\vert_{J=0}=\frac{\delta^n \mS_{(0)}[h_{\mu\nu}(J)]}{\delta J^n}\Big\vert_{J=0}=\sum_i(-2w_i\varphi)^n\Lag_{(0),i}
\ee
so that the resummed Lagrangian for the term of degree $w_i$ is
\be
\Lag_{(i)}=\sum_{n=0}^\infty \frac{\delta^n \mS[h_{\mu\nu}(J)]}{\delta J^n}\Big\vert_{J=0}=\left[\sum_{n=0}^\infty\left(-2w_ i\varphi\right)^n\right]\Lag_{(0),i}=\frac{1}{1+2w_i\varphi}\Lag_{(0),i}.
\ee
This exactly reproduces the results of Sec. \ref{sec:Conformalscalar} since the kinetic term of the scalar has degree $1$ while any potential term for a scalar field has degree $2$. Notice that for a term of zero degree there is no correction, in accordance with the fact that the energy-momentum tensor of a zeroth weight (i.e. conformally invariant) is traceless.

After warming up with the simplest example (which in fact allows for a full resolution of the problem) we can proceed to develop a general framework. Let us consider a leading order interaction to the energy-momentum tensor of the general form
\be
\Lag_{\rm}^{(1)}=\Omega_{\mu\nu} T^{\mu\nu}
\ee
where $\Omega_{\mu\nu}$ is some rank-2 tensor built out of the field which we want to couple to $T^{\mu\nu}$, be it the scalar or the vector under consideration throughout this work. Then, it is easy to see that the generating effective metric will be given by
\be
h_{\mu\nu}=\exp\big[-2J\omega_{\mu\nu}{}^{\alpha\beta}\big]\eta_{\alpha\beta}
\label{eq:defheff}
\ee
with $\omega_{\mu\nu}{}^{\alpha\beta}$ some rank-4 tensor satisfying $\omega_{\mu\nu}{}^{\alpha\beta}\eta_{\alpha\beta}=\Omega_{\mu\nu}$. This effective metric indeed generates the desired interaction from the functional derivative
\be
\Lag_{\rm}^{(1)}=\frac{\delta \mS_{(0)}[h_{\mu\nu}(J)]}{\delta J}\Big\vert_{J=0},
\ee
while the higher order interactions are simply
\be
\Lag_{\rm}^{(n)}=\frac{\delta \mS_{(n-1)}[h_{\mu\nu}(J)]}{\delta J}\Big\vert_{J=0}.
\ee
This general construction can be straightforwardly applied to the cases that we have considered in the precedent sections. The coupling of the scalar field corresponds to taking $\omega_{\mu\nu}{}^{\alpha\beta}$ proportional to the identity in the space of rank-4 tensors. The algebraic vector field coupling corresponds to $\omega_{\mu\nu}{}^{\alpha\beta}=\frac14\(b_1A_\mu A_\nu+b_2A^2\eta_{\mu\nu}\) \eta^{\alpha\beta}$, while the gauge invariant coupling is generated by $\omega_{\mu\nu}{}^{\alpha\beta}=b_1 F_\mu{}^\alpha F_\nu{}^\beta+\frac14 b_2 F^2\eta_{\mu\nu}\eta^{\alpha\beta}$.

An interesting case is when $\Omega_{\mu\nu}$ does not depend on $\eta_{\mu\nu}$, which happens when $\omega_{\mu\nu}{}^{\alpha \beta}$ is linear in the inverse metric. To see why this is particularly interesting, let us obtain the second correction to the original action, that will be given, in general, by
\bea
\Lag^{(2)}&=&\left(\frac{\delta \mS_{(1)}[h_{\mu\nu}]}{\delta J}\right)_{J=0}=
\left[\frac{\delta}{\delta J}\int\d^4x\left(\frac{\delta \mS_{(0)}[h_{\mu\nu}]}{\delta J}\right)_{J=0}\right]_{J=0}\\
&=&\left[\frac{\delta}{\delta J}\int\d^4x\left(-2\omega_{\mu\nu}{}^{\alpha\beta}\eta_{\alpha\beta}\frac{\delta \mS_{(0)}[h_{\mu\nu}]}{\delta h_{\mu\nu}}\Big\vert_{J=0}\right)\right]_{J=0}\nonumber
=-2\left[\frac{\delta}{\delta J}\int\d^4x\omega_{\mu\nu}{}^{\alpha\beta}(J)h_{\alpha\beta}(J)\frac{\delta \mS_{(0)}[h_{\mu\nu}]}{\delta h_{\mu\nu}}\right]_{J=0}.\nonumber
\eea
where we have used the chain rule and the definition of the generating metric (\ref{eq:defheff}). Now, the last expression is substantially simplified if $\Omega_{\mu\nu}=\omega_{\mu\nu}{}^{\alpha\beta} \eta_{\alpha\beta}$ does not depend on the metric $\eta_{\mu\nu}$ because, in that case, $\omega_{\mu\nu}{}^{\alpha\beta} \eta_{\alpha\beta}$ will not acquire a dependence on $J$ so we obtain
\be
\Lag^{(2)}=(-2)^2\omega_{\mu\nu}{}^{\alpha\beta} \eta_{\alpha\beta} \omega_{\rho\sigma}{}^{\gamma\delta} \eta_{\gamma\delta}\left(\frac{\delta^2 \mS_{(0)}[h(J)]}{\delta h_{\mu\nu}\delta h_{\rho\sigma}}\right)_{J=0}.
\ee
It is then straightforward to iterate this process and arrive at the following expression for the final Lagrangian
\be
\Lag_{\rm}=\sum_{n=0}^{\infty}(-2)^n\frac{\delta^n \mS_{(0)}[h(J)]}{\delta h_{\mu_1\nu_1}\cdots \delta h_{\mu_n\nu_n}}\Big\vert_{J=0}\Omega_{\mu_1\nu_1}\cdots \Omega_{\mu_n\nu_n}.
\ee
This (asymptotic) expansion reminds of a Taylor expansion, though without the required $1/n!$. It is not difficult to motivate the appearance of the missing factorial by simply imposing that the variation should appear at the level of the field equations instead of the action, similarly to what we did for the scalar field case at the end of Sec. \ref{sec:Conformalscalar}. In that case, the resulting series will be
\be
\Lag_{\rm}=\sum_{n=0}^{\infty}\frac{(-2)^n}{n!}\frac{\delta^n \mS_{(0)}[h]}{\delta h_{\mu_1\nu_1}\cdots \delta h_{\mu_n\nu_n}}\Omega_{\mu_1\nu_1}\cdots \Omega_{\mu_n\nu_n}.
\label{eq:TotalLeffmetric}
\ee
which admits a straightforward resummation as the Taylor expansion of the seed action $\mS_{(0)}$ around $2\Omega_{\mu\nu}$, i.e., the total action will be $\mS=\mS_0[\eta_{\mu\nu}-2\Omega_{\mu\nu}]$. Of course, this reproduces the well-known result for gravity when $\Omega_{\mu\nu}$ is identified with the metric perturbation $-2h_{\mu\nu}$. We see here that the same applies when a scalar field is derivatively coupled to matter fields as $\partial_\mu\varphi\partial_\nu\varphi T^{\mu\nu}$ or a vector field is coupled as $A_\mu A_\nu T^{\mu\nu}$. In both cases, the resulting coupling to matter fields is through a disformal metric $g_{\mu\nu}=\eta_{\mu\nu}-2A_\mu A_\nu$ and $g_{\mu\nu}=\eta_{\mu\nu}-2\partial_\mu \varphi \partial_\nu\varphi$ respectively. This result shows that the couplings of the precedent sections with $b_2=0$ are indeed special because they satisfy the above condition of having the corresponding $\Omega_{\mu\nu}$ independent of the metric and, therefore, admit a resummation in terms of an effective metric. 

Finally, let us notice that the expression (\ref{eq:TotalLeffmetric}) also serves as a starting point to explore a class of "non-metric" theories, i.e., theories where the coupling is not through an effective metric, analogous to the theories explored in \cite{Blanchet:1992rx} as non-metric departures from GR.

\section{Phenomenology}
\label{Sec:Phenomenology}
The phenomenology associated with the different scalar and vector couplings discussed along this work is very rich and depends on the particular term. The best known case corresponds to the scalar coupling to the trace of the energy momentum tensor. These types of couplings leads to the standard Jordan-Brans-Dicke framework and dilaton structures. In this section, we would like to summarize the main phenomenology related to disformal scalars and vectors. In particular, we would like to focus on the phenomenology that is shared by both fields due to the equivalence theorem. For instance, for the vectorial case, we will focus in the regime where its mass is negligible with respect to the physical energy involved in the process. Indeed, following an effective theory approach, we can work in the perturbation limit. In such a case, the dimension 6 interaction term can be written as:
\begin{equation}
\mathcal{L}_V = \frac{1}{\Mvc^2}A_\mu A_\nu T^{\mu\nu}\;.
\label{eq:vcoupling}
\end{equation}
This term dominates the distinctive phenomenology associated to the vector fields discussed along this work. This type of interaction corresponds also to the vector modes associated to the metric in theories with additional spatial dimensions \cite{DoMa}. The phenomenology of these {\it graviphotons} have been studied for the brane world scenario in different works under the name of {\it brane vectors} \cite{Clark,Clark:2008fn,Clark:2008zw}. In this framework, there is a Higg-like mechanism that provides mass to the graviphotons. At high energies with respect to such a mass, the longitudinal mode of these vectors can be identified with the branons $\varphi$, the scalar degree of freedom associated to the fluctuation of the brane along the extra dimensions \cite{DoMa}. Within this regime, the experimental signatures of these vectors can be computed by the branon disformal coupling \cite{Cembranos:2016jun}:
\begin{equation}
\mathcal{L}_D = \frac{1}{m^2\,\Mvc^2}\partial_\mu \varphi \partial_\nu \varphi T^{\mu\nu}\;,
\label{eq:coupling}
\end{equation}
where $m$ is the mass of the vector field. A detailed analysis of the constraints to the above interaction has been developed in the context of branon fields \cite{Sundrum,DoMa,BSky,Alcaraz:2002iu,thesis}. This term dominates the distinctive observational signatures of disformal scalars. Not only potential searches in colliders have been studied in different works \cite{strumia,Alcaraz:2002iu}, but also astrophysical constraints have been analyzed, such as the ones coming from cooling of stelar objects \cite{Kugo,CDM,CDM2} or the associated to the relic abundances of this type of massive vectors \cite{Clark,Clark:2008fn}.

At high energies, the vector fields can be searched by analyzing Lagrangian (\ref{eq:coupling}). In this case, the vector mass times the vector coupling is the combination of parameters which suffers the constraints from present data. On the contrary, for the disformal scalars, the constraints apply directly to the disformal scalar coupling. They may be detected at the Large Hadron Collider (LHC) or in a future generation of accelerators
\cite{Alcaraz:2002iu,Brax:2012ie,Brax:2014vva,Clark,Coll,L3,Cembranos:2004jp,LHCDirect,Landsberg:2015pka,Khachatryan:2014rwa,BWRad}.
For the case of the LHC, the most sensitive production process is gluon fusion giving a gluon in addition to
a pair of longitudinal vectors or disformal scalars; and the quark-gluon interaction giving a quark and a pair of the commented particles. These processes contribute to the
monojet $J$ plus transverse missing momentum and energy signal.
An additional process is the quark-antiquark annihilation giving a photon and the mentioned pair of new modes.
In such a case, the signal is a single photon in addition to the transverse missing momentum and energy.
The cross-section of the subprocesses were computed in Refs. \cite{Cembranos:2004jp} and \cite{LHCDirect}.
The analysis of the single photon channel is simpler and cleaner but the monojet channel is more sensitive.

In addition to these processes, there are other complementarity constraints on the same combination of parameters
corresponding to other collider data. A summary of
these analyses can be found in Table \ref{tabHad} \cite{DoMa,BW2,Clark,Coll,Cembranos:2004jp,LHCDirect}.
In this Table, the limits coming from HERA, LEP-II and Tevatron
are compared with the present restrictions from LHC running at a centre of mass energy
(c.m.e.) of 8 TeV and the prospects for the LHC running at 14 TeV c.m.e.
with full luminosity. Other missing transverse momentum and energy
processes, such as those related to the monolepton channel \cite{Brax:2014vva}, are
also potential signatures of the models developed in this work. 
In the same reference, the authors
discuss other different phenomenological signatures, but they are subdominant due to the
important dependence of the interaction with the energy.

On the other hand, it has been shown that the new modes under study introduce radiative corrections, which generate new couplings among SM particles, which can be described by an
effective Lagrangian. Although the study of such processes demands the introduction of new parameters, they can provide interesting effects
in electroweak precision observables, anomalous magnetic moments, or SM four particle interactions  \cite{BWRad}.

\begin{table}
\begin{center}{
\begin{tabular}{||c|ccc||}
\hline Experiment
&
$\sqrt{s}$(TeV)& ${\mathcal L}$(pb$^{-1}$)&$\sqrt{m\Mvc}$(GeV)\\
\hline
%
%
HERA$^{\,1}
$& 0.3 & 110 &  19
\\
Tevatron-II$^{\,1}
$& 2.0 & $10^3$ &  304
\\
Tevatron-II$^{\,2}
$& 2.0 & $10^3$ &   285
\\
LEP-II$^{\,2}
$& 0.2 & 600 &  214
\\
%
%
%
LHC$^{\,2}
$& 8 & $19.6\times10^{3}$ &   523
\\
\hline
LHC$^{\,1}
$& 14 & $10^5$ &  1278
\\
LHC$^{\,2}
$& 14 & $10^5$ &   948
\\
\hline
\end{tabular}
}
\caption{
Summary table for the phenomenology of vectors and disformal scalars coupled
to the energy-momentum tensor at colliders.
Monojet and single photon analyses are labeled by the upper indices ${1,2}$, respectively.
Present bounds and prospects for the LHC \cite{Landsberg:2015pka,Khachatryan:2014rwa,Cembranos:2004jp,LHCDirect}
are compared with constraints from LEP \cite{Alcaraz:2002iu,L3}, HERA and Tevatron \cite{Cembranos:2004jp}.
$\sqrt{s}$ means the centre of mass energy associated with the total process;
${\mathcal L}$ denotes the total integrated luminosity;
 $\sqrt{m\Mvc}$ is the constraint at the $95\;\%$ confidence level by assuming a very light vector
 (in the limit $m\rightarrow0$). The effective coupling is not valid for energy scales
$\Lambda^2\gtrsim 8\pi\sqrt{2} m \Mvc$ \cite{BWRad}.
}

\label{tabHad}\end{center}
\end{table}

We can also compute the thermal relic abundance corresponding to these new fields by assuming they are stable \cite{CDM,CDM2,Clark:2008fn}. The larger the coupling scale $M$, the weaker the annihilating cross-section into SM particle-antiparticle pairs, and the larger the relic abundance. This is the expected conclusion since the sooner the decoupling occurs, the larger the relic abundance is. Therefore the cosmological restrictions related to the relic abundance are complementary to those coming from particle accelerators. Indeed, a constraint such as $\Omega_{D} < \Od(0.1)$ means a lower limit for the value of the cross-sections in contrast with the upper limits commented above from non observation at colliders. If we assume that the DM halo of the Milky Way has an important amount of these new vectors or scalars,
its flux on the Earth could be sufficiently large to be measured in direct detection experiments. These experiments measure the rate $R$, and energy
$E_{R}$ of nuclear recoils. 
These constraints depend on different astrophysical assumptions as it has been discussed in different analyses \cite{CDM,CDM2,Clark:2008fn,Cembranos:2016jun,LHCDirect}.

If the abundance of these new particles is significant, they cannot only be detected by direct detection experiments, but also by indirect ones. In fact, a pair of vectors or scalars can annihilate into ordinary particles such as leptons, gauge bosons, quarks or Higgs bosons. Their annihilations from different astrophysical regions produce fluxes of cosmic rays. Depending on the characteristics of these fluxes, they may be discriminated from the background. After the annihilation and propagation, the particle species that can be potentially detected by different detectors are gamma rays, neutrinos and antimatter (fundamentally, positrons and antiprotons). In particular, gamma rays and neutrinos have the advantage of maintaining their original trajectory. Indeed, this analysis are more sensitive for the detection of these signatures \cite{vivi}.

\begin{figure*}[htbp]
\centering
\includegraphics[width=12cm]{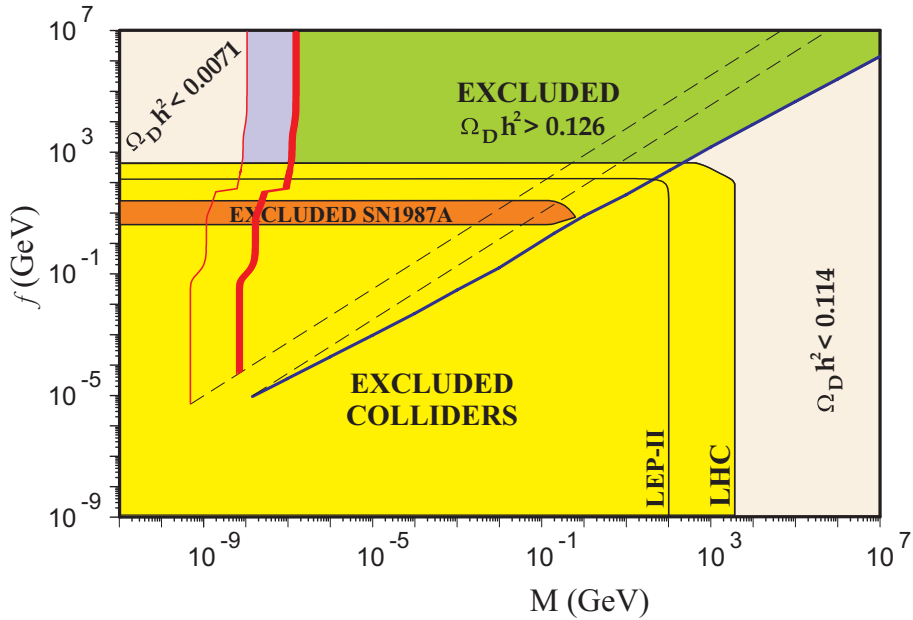}
\caption{In this figure we show the combined exclusion plot for a model with a single disformal scalar described by the branon
Lagrangian. $M$ is the scalar (branon) mass, whereas $f$ is the brane
tension scale (which verifies $2 f^4=M_{\text{sd}}^4$).
Within the (green) upper area, the model is ruled out because these scalars are overproduced, The (purple) left 
region is excluded by structure formation, whereas the (yellow) lower
regions are excluded by LEP-II and LHC single photon analysis. Supernovae cooling rules out the horizontal (orange) area. The (blue) solid line on the right is associated with a cold DM behaviour with the proper relic density. The two (red) solid lines on the left are related to a hot DM behaviour. This figure is taking from Ref. \cite{Cembranos:2016jun}, where a more detailed explanation can be found.}
\label{FigBranons}
\end{figure*}

On the other hand, there are astrophysical observations that are able to constraint the parameter space of the new fields studied in this work independently of their abundance. For example, one of the most successful predictions of the standard cosmological model is the relative abundances of primordial elements. These abundances are sensitive to several cosmological parameters and were used in Refs. \cite{CDM} and \cite{CDM2}
in order to constrain the number of light fields. These restrictions apply in this case since the new particles will behave as dark radiation for small enough masses. For instance, the production of $^4$He increases with an increasing rate of the expansion $H$ and the Hubble parameter depends on the total amount of radiation. However, these restrictions are typically important for relatively strong couplings. On the contrary, if the new modes decouple above the QCD phase transition, $ m \Mvc \sim 10000 $ GeV$^2$, the limit increases so much that the restrictions become extremely weak \cite{CDM2}. Different astrophysical bounds can be obtained from modifications of cooling processes in stellar objects like supernovae \cite{Kugo,CDM,CDM2,Brax:2014vva}. These processes take place by energy loosing through light particles such as photons and neutrinos. However, if the mass of the new particles is low enough, these new particles are expected to carry a fraction of this energy,
depending on their mass and the coupling to the SM fields. These constraints are restricting up to masses of order of the GeV. For heavier fields, the limits on the coupling disappear due to the short value of the mean free path of the vector particle inside the stellar object \cite{CDM2}. A summery of these astrophysical and cosmological constraints for a particular model of a single disformal scalar described by the branon Lagrangian can be seen in Fig. \ref{FigBranons}.

In the previous paragraphs, we have summarized the phenomenology associated to the abundance of these new vectors and disformal scalars by assuming it was generated by the thermal decoupling process in an expanding universe. However, if the reheating temperature $T_{RH}$ after inflation was sufficiently low, then these new fields were never in thermal equilibrium with the primordial plasma \cite{nonthermal}. However, still there is
the possibility for them to be produced non-thermally, very much in the same ways as axions \cite{axions}
or other bosonic degrees of freedom \cite{Cembranos:2012kk,santos}. This possibility modifies some of the previous astrophysical signatures, since they have typically associated a much lower mass. In particular, the potential isotropies related to the coherent relic density of the new vectors could constitute a very distinctive signature \cite{santos}.

\section{Discussion}
\label{Sec:Discussion}
In this work we have explored the construction of theories involving scalar and vector fields following a procedure inspired by gravity so that the fields are coupled to the total energy-momentum tensor. This defining property is not free from ambiguities arising from the different definitions of the energy-momentum tensor that we fix by using the Hilbert prescription. Another ambiguity present in the construction comes from requiring a coupling at the level of the action or a coupling such that it is the source of field equations. We have mostly employed the former, although we have commented upon this subject and considered the latter in some specific cases.

We have started by reviewing some known results that exist in the literature concerning scalar fields. We have then extended these results to theories where the coupling of the scalar to the energy-momentum tensor respects a shift symmetry, i.e., the scalar couples derivatively. We have reduced the resummation of the infinite perturbative series to the resolution of some differential equations, whose properties we explain in some detail in Appendix \ref{Appendix}. We have nevertheless highlighted some particularly remarkable parameters choices where the solutions can be readily obtained as well as the existence of polynomial solutions. The difficulty encountered in obtaining the full theory within the second order formalism encouraged us to consider the problem from a first order perspective, where it is known that the self-coupling procedure simplifies substantially. We have indeed confirmed that this is also the case for the problem at hand and we managed to reduce the resumation to the resolution of algebraic instead of differential equations. We ended our tour on scalar fields by exploring couplings to matter fields taking a scalar as a proxy.

After considering scalar fields, we delved into vector fields coupled to the energy-momentum tensor. This case offers a richer spectrum of results owed to the possibility of having a gauge symmetry. We have first constructed the theory for a vector coupled to the energy-momentum tensor without taking care of the gauge invariance. We have then considered gauge-invariant couplings where the $U(1)$ symmetry is realised in the usual way in the vector field so that the couplings are through the field strength of the vector. We have however commented on the interesting possibility of having the $U(1)$ symmetry non-linearly realised in the vector field so that the resulting action could explicitly contain the vector not necessarily through $F_{\mu\nu}$, but still keeping two propagating dof's for the vector. This path would be interesting to pursue further, specially if it can be linked to the conservation of the energy-momentum tensor (or some related conserved current) so that the theories are closer to the gravity case, i.e., the consistent couplings are imposed by symmetries rather than by a somewhat ad-hoc prescription. Similarly to the scalar field theories, the resummed actions for the vector fields reduced to solving some differential equations with similar features. However, unlike for theories with scalars, the first order formalism applied to the vector field case did not result in any simplification of the iterative problem. Finally, we also considered couplings to a scalar as a proxy for the matter fields. Although we did not consider other types of fields throughout this work, it would not be difficult to include higher spin fields. 

After constructing the actions for our self-interacting fields through the energy-momentum tensor, we have discussed the impact of superpotential terms arising from the ambiguity in the energy-momentum tensor. We have shown that these superpotential terms lead to the generation of Galileon interactions for both, the scalar and the vector. 

Given the motivation from gravity theories that triggered our study and the prescribed couplings to the energy-momentum tensor, we have explored generating functionals defined in terms of an effective metric. This method allowed us to re-derive in a simpler way some of the results in the previous sections and give a general procedure to generate all the interactions. Moreover, we have shown that the generating functional procedure greatly simplifies if the leading order correction to the energy-momentum tensor does not depend on the metric. If that is the case, the series can be regarded as a Taylor expansion and the final resummed action is simply the original action coupled to an effective metric.

Finally, we have considered the phenomenology of these type of theories. The possible experimental signatures associated with the
scalar and vector couplings discussed along this work is very rich and depends on the particular term under study. We have focused on
the phenomenology related to disformal scalars and vectors. In fact, at high energies, the longitudinally polarized vector and the 
disformal scalar are related by the equivalence theorem. Following an effective theory approach, we can work in the perturbation limit.
In such a case, monojet and single photon analysis at the LHC are the most sensitive signatures, constraining the dimensional couplings
at the TeV scale. On the other hand, both models can support 
viable dark matter candidates, offering a broad range of
astrophysical and cosmological observable possibilities depending
of their stability. Note that given the quadratic character of the leading order interaction for the derivative couplings of the scalar or the non-gauge invariant couplings of the vector, the corresponding force will decay faster than the Newtonian behaviour $1/r^2$. However, if there is some vev for the fields, be it $\langle \partial_\mu\varphi\rangle$ or $\langle A_\mu\rangle$ for the scalar and the vector respectively, then we can recover the Newtonian-like force with a coupling constant determined by the vev of the field. Indeed the non-linear nature of the interactions may give rise to screening mechanisms, with interesting phenomenology for dark energy
models.

An interesting question that we have not addressed in this work is the relation of our resulting actions with geometrical frameworks. We have already given some arguments in Sec. \ref{Sec:EffectiveMetrics} as to what extent our constructed theories could be regarded as couplings to an effective metric. For the massless scalar field case, it is known that the resulting action can be interpreted in terms of the Ricci scalar of a metric conformally related to the Minkowski metric (which is nothing but Nordstr\o m's theory). It would be interesting to check if analogous results exist for the derivatively coupled case and/or the vector field case, for instance in terms of curvature objects of disformal metrics. Perhaps a suited framework for these theories would be the arena provided by Weyl geometries or generalised Weyl geometries which naturally contain an additional vector field, which could also be reduced to the gradient of a scalar, in which case it is known as Weyl integrable spacetimes or WIST. Another extension of our work can be associated with the analysis of multi-field theories and the study of the role played by internal symmetries 
within these constructions.

\acknowledgments 
We thank Pepe Aranda for useful discussions. This work is supported by the MINECO (Spain) projects FIS2014-52837-P and FIS2016-78859-P (AEI/FEDER). JBJ acknowledges the support of the Spanish MINECO’s “Centro de Excelencia Severo Ochoa” Programme under grant SEV-2016-0597.
 JMSV acknowledges the support of Universidad Complutense de Madrid through the predoctoral grant CT27/16. 

\appendix
\section{Discussion of the recurrent differential equations}
\label{Appendix}
In this appendix we will discuss the differential equation that we have recurrently obtained throughout this paper. The equation can be expressed as
\be
2x^2F'(x)+(px-1)F(x)+1=0
\label{Eq:master}
\ee
with $p$ a constant. Taking for instance our first encounter with this type of equation in (\ref{Eq:eqsderivativecoupling}), that we reproduce here for convenience
\bea
\mK&=&1+(b_1-2b_2)X\mK+2(b_1+b_2)X^2\mK',\\
\mU&=&1-(b_1+4b_2)X\mU+2(b_1+b_2)X^2\mU',
\label{Eq:eqsderivativecoupling2}
\eea
it is easy to check that they can be brought into the form (\ref{Eq:master}) by simply rescaling $X\to X/(b_1+b_2)$, which is always legitimate except for $(b_1+b_2)=0$, in which case, as we already discussed, the equations become algebraic. The values for the constant parameter $p$ are then $p=(b_1-2b_2)/(b_1+b_2)$ and $p=-(b_1+4b_2)/(b_1+b_2)$ for $\mK$ and $\mU$ respectively. It is clear that $x=0$ is a singular point of the equation. If we evaluate the equation at $x=0$ we find $F(0)=1$, as it should since the series of the functions satisfying this equation start at 1. Furthermore, by taking subsequent derivatives of the equation and evaluating at $x=0$, we can reproduce the corresponding perturbative series, which is in turn an asymptotic expansion. Alternatively, we can seek for solutions in the form of a power series $F=\sum_{n=0}F_nx^n$. By substituting this series in the equation we find that $F_0=1$ and the following recurrent formula for the coefficients with $n\geq1$:
\be
F_n=\Big[p+2(n-1)\Big]F_{n-1}, \quad n\geq1, 
\label{Eq:recursive}
\ee
which can be readily solved as
\be
F_n=\prod_{j=1}^n\Big[p+2(n-j)\Big],\quad n\geq1.
\label{Eq:Fn}
\ee
From this general solution we see that $F_1=p$ so that if we choose the parameters such that $p=0$, then the solution is simply $F=1$ because then only $F_0$ remains non-vanishing. This result of course corresponds to the choice of parameters that cancelled either the corrections to the kinetic terms or to the potential discussed throughout this work. In particular, we see that $p=0$ corresponds to $b_1=2b_2$ and $b_1=-4b_4$ in (\ref{Eq:eqsderivativecoupling2}), as we already found in Sec. \ref{Sec:ScalarDerivative}.

The fact that we can write the general solution in terms of a recursive relation is due to the iterative procedure prescribed to built our actions and this in turn has noteworthy consequence, namely, we can choose parameters such that the solution becomes polynomial. Since the coefficients $F_n$ are recursively given by (\ref{Eq:recursive}), if we have that $F_{r+1}=0$ for some $r$, then $F_{n}=0$ for $n>r$ and, thus, the solution is a polynomial of degree $r$. The result commented in the previous paragraph that $F=1$ for $p=0$ is a particular case of this general result with $r=0$. In general, if we want the solution to be a polynomial of degree $r$ we will need to impose $p+2r=0$. As an illustrative example, if we want $\mK$ and $\mU$ in (\ref{Eq:eqsderivativecoupling2}) to be polynomials of degree $r$ and $s$ respectively, we need to choose the parameters $b_1$ and $b_2$ satisfying
\be
\frac{b_1-2b_2}{b_1+b_2}=-2r,\quad\quad\frac{b_1+4b_2}{b_1+b_2}=2s.
\ee
These equations do not admit a general solution for arbitrary $r$ and $s$. If we eliminate for instance $b_1$, we end up with the equation $b_2(1+r-s)=0$ (for $r$ and $s$ integers), that imposes the relation $s=1+r$ and leaves $b_2$ as a free parameter. The solution for $b_1$ is then $b_1=2b_2(1-r)/(1+2r)$. Thus, in this particular case, if we want both functions to become polynomials, they cannot be of arbitrary degree, but $\mU$ must be one degree higher than $\mK$. For $r=0$, we have $s=1$ and the corresponding theory has $b_1=2b_2$, i.e., for those parameters $\mK$ is constant and $\mU$ is polynomial of degree 1, as we found in Sec. \ref{Sec:ScalarDerivative}.

If we do not impose the solutions to be polynomials, we can do better than the solution expressed as the (asymptotic) series with general coefficients given in (\ref{Eq:Fn}) and obtain the general solution of the equation in terms of known functions, which can be expressed as
\be
F(x)=e^{-1/(2x)}\vert x\vert ^{-p}\left[C-\frac{1}{2}\int e^{1/(2x)}\vert x\vert ^{p-2}\d x\right]
\ee
with $C$ a constant that must be chosen in order to have a well-defined solution at $x=0$. The above solution can also be expressed in terms of exponential integral functions as follows
\be
F(x)=C\vert x\vert^{-p/2}e^{-1/(2x)}-\frac{1}{2x}e^{-1/(2x)}\Ei_{p/2}(-1/(2x)).
\ee
Since the exponential integral has the asymptotic expansion $\Ei_n(x)\sim e^{-x}/x$ for large $x$, the second term in the above solution is regular at the origin and, thus, in order to have a regular solution we must impose $C=0$ so that the desired solution is finally
\be
F(x)=-\frac{e^{-1/(2x)}}{2x}\Ei_{p/2}(-1/(2x)).
\ee
The exponential integral presents a branch cut for the negative axis. In the final solution, it must be interpreted as the real part of the analytic continuation to the complex plane.

\begin{figure*}[htbp]
\centering
\includegraphics[width=8cm]{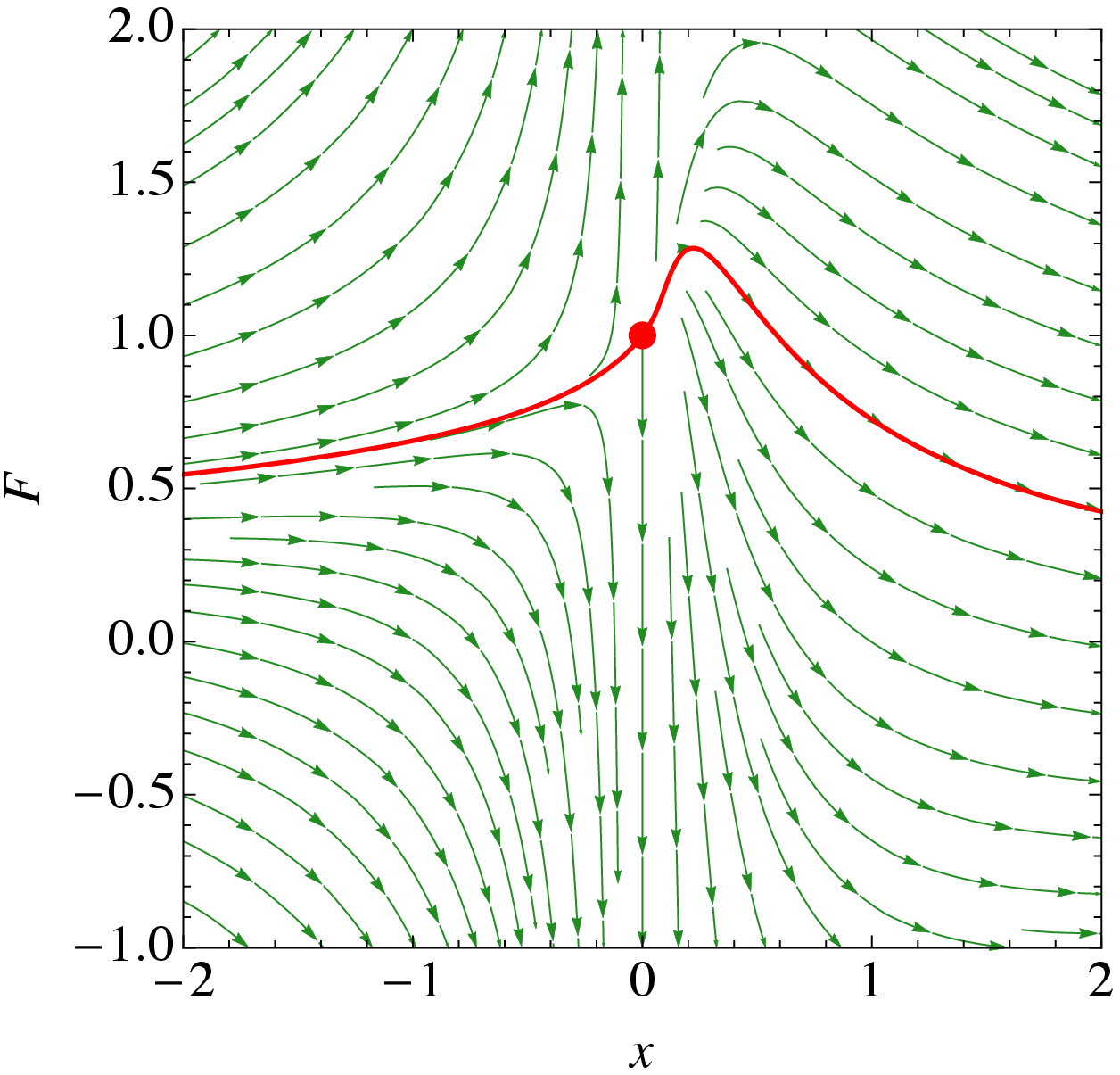}\hspace{0.5cm}
\includegraphics[width=8cm]{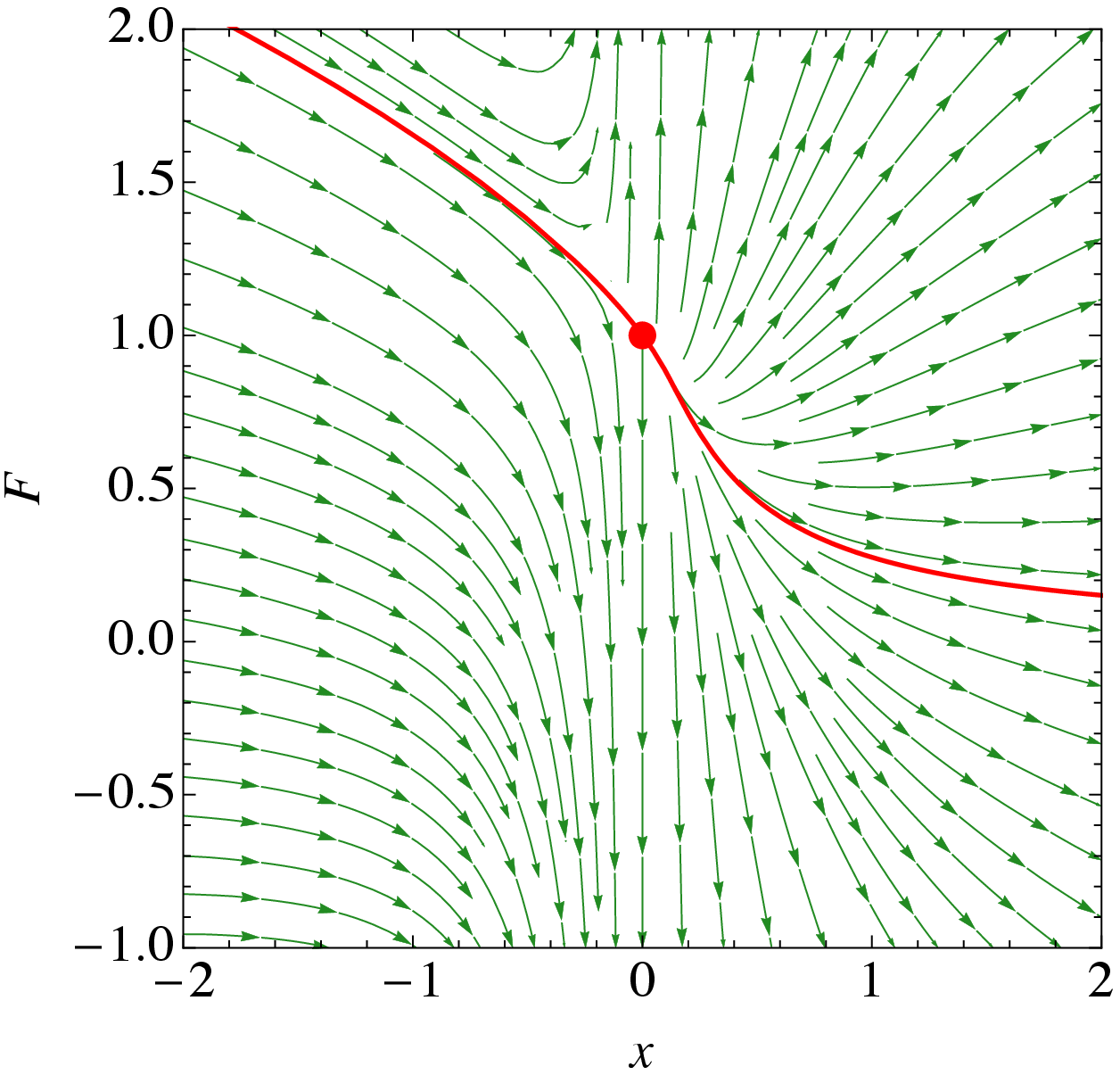}
\caption{In these plots we show the phase maps corresponding to $p=1$ (left) and $p=-1$ (right) of the associated autonomous system. We also highlight the unique solution that smoothly goes through the critical point $(x=0,F=1)$ (red dot) from the left. Notice however that the continuation in the right is not uniquely determined due to the existence of a centre manifold that can be clearly appreciated in the plots. We have simply singled out one of them.}
\label{FigA}
\end{figure*}

We can gain some more insights on the equation by considering the equivalent autonomous system
\bea
\dot{F}(t)&=&\Big(1-px(t)\Big)F(t)-1\nonumber\\
\dot{x}(t)&=&2x^2(t)
\eea
where the dot means derivative with respect to $t$. The integral curves of this autonomous system are the solutions of our original equation. The only fix point is given by $x=0$, $F=1$ and the associated eigenvalues are $0$ and $1$. Moreover, it is also easy to see that the $x$-axis is indeed a separatrix, ($\dot{x}=0$ on that axis) so the only solution that can cross it must necessarily pass through $(F=1,x=0)$. The fact that there is a negative eigenvalue indicates the existence of a centre (slow) manifold. We will not provide a detailed analysis of the general properties of the associated autonomous system, but we will simply enumerate some important properties that are illustrated in Fig. \ref{FigA}. We can see that there is only one trajectory than can smoothly pass through $x=0$ so the requirement to have a well-defined behaviour for $x\to0$ singles out one solution on the left half plane. However, since the trajectories approach $x=1$ along a centre manifold from the right half-plane, one can, in principle, match the regular solution coming from the left to any of the existing solutions in the right half-plane.


\end{document}